\input harvmac.tex

% \draft

\lref\srednicki{T.~Falk, R.~Rangarajan and M.~Srednicki,
%``Dependence of density perturbations on the coupling constant in a simple model of inflation,''
Phys.\ Rev.\ D {\bf 46}, 4232 (1992)
[arXiv:astro-ph/9208002].
%%CITATION = ASTRO-PH 9208002;%%
}

\lref\srednickivariance{
M.~Srednicki,
%``Cosmic variance of the three point correlation function of the cosmic microwave background,''
arXiv:astro-ph/9306012.
%%CITATION = ASTRO-PH 9306012;%%
}

\lref\jm{
J.~M.~Maldacena,
%``The large N limit of superconformal field theories and supergravity,''
Adv.\ Theor.\ Math.\ Phys.\  {\bf 2}, 231 (1998)
[Int.\ J.\ Theor.\ Phys.\  {\bf 38}, 1113 (1999)]
[arXiv:hep-th/9711200].
%%CITATION = HEP-TH 9711200;%%
}

\lref\wittenhol{
E.~Witten,
%``Anti-de Sitter space and holography,''
Adv.\ Theor.\ Math.\ Phys.\  {\bf 2}, 253 (1998)
[arXiv:hep-th/9802150].
%%CITATION = HEP-TH 9802150;%%
}

\lref\gkp{
S.~S.~Gubser, I.~R.~Klebanov and A.~M.~Polyakov,
%``Gauge theory correlators from non-critical string theory,''
Phys.\ Lett.\ B {\bf 428}, 105 (1998)
[arXiv:hep-th/9802109].
%%CITATION = HEP-TH 9802109;%%
}

\lref\wise{
T.~J.~Allen, B.~Grinstein and M.~B.~Wise,
%``Nongaussian Density Perturbations In Inflationary Cosmologies,''
Phys.\ Lett.\ B {\bf 197}, 66 (1987).
%%CITATION = PHLTA,B197,66;%%
}

\lref\gangui{A.~Gangui, F.~Lucchin, S.~Matarrese and S.~Mollerach,
%``The Three point correlation function of the cosmic microwave background in inflationary models,''
Astrophys.\ J.\  {\bf 430}, 447 (1994)
[arXiv:astro-ph/9312033].
%%CITATION = ASTRO-PH 9312033;%%
}

\lref\acquaviva{V.~Acquaviva, N.~Bartolo, S.~Matarrese and A.~Riotto,
%``Second-order cosmological perturbations from inflation,''
arXiv:astro-ph/0209156, revised version.
%%CITATION = ASTRO-PH 0209156;%%
}

\lref\starobinsky{
A.~A.~Starobinsky,
%``Dynamics Of Phase Transition In The New Inflationary Universe Scenario And Generation Of Perturbations,''
Phys.\ Lett.\ B {\bf 117}, 175 (1982).
%%CITATION = PHLTA,B117,175;%%
}

\lref\hawking{
S.~W.~Hawking,
%``The Development Of Irregularities In A Single Bubble Inflationary Universe,''
Phys.\ Lett.\ B {\bf 115}, 295 (1982).
%%CITATION = PHLTA,B115,295;%%
}

\lref\guth{
A.~H.~Guth and S.~Y.~Pi,
%``Fluctuations In The New Inflationary Universe,''
Phys.\ Rev.\ Lett.\  {\bf 49}, 1110 (1982).
%%CITATION = PRLTA,49,1110;%%
}
\lref\porrati{
L.~Girardello, M.~Petrini, M.~Porrati and A.~Zaffaroni,
%``Novel local CFT and exact results on perturbations of N = 4 super  Yang-Mills from AdS dynamics,''
JHEP {\bf 9812}, 022 (1998)
[arXiv:hep-th/9810126].
%%CITATION = HEP-TH 9810126;%%
}

\lref\gubserctheorem{
D.~Z.~Freedman, S.~S.~Gubser, K.~Pilch and N.~P.~Warner,
%``Renormalization group flows from holography supersymmetry and a  c-theorem,''
Adv.\ Theor.\ Math.\ Phys.\  {\bf 3}, 363 (1999)
[arXiv:hep-th/9904017].
%%CITATION = HEP-TH 9904017;%%
}

\lref\bst{
J.~M.~Bardeen, P.~J.~Steinhardt and M.~S.~Turner,
%``Spontaneous Creation Of Almost Scale - Free Density Perturbations In An Inflationary Universe,''
Phys.\ Rev.\ D {\bf 28}, 679 (1983).
%%CITATION = PHRVA,D28,679;%%
}

\lref\freedman{
O.~DeWolfe and D.~Z.~Freedman,
%``Notes on fluctuations and correlation functions in holographic  renormalization group flows,''
arXiv:hep-th/0002226.
%%CITATION = HEP-TH 0002226;%%
}

\lref\frolov{
G.~Arutyunov, S.~Frolov and S.~Theisen,
%``A note on gravity-scalar fluctuations in holographic RG flow  geometries,''
Phys.\ Lett.\ B {\bf 484}, 295 (2000)
[arXiv:hep-th/0003116].
%%CITATION = HEP-TH 0003116;%%
}

\lref\skenderisone{
M.~Bianchi, D.~Z.~Freedman and K.~Skenderis,
%``How to go with an RG flow,''
JHEP {\bf 0108}, 041 (2001)
[arXiv:hep-th/0105276].
%%CITATION = HEP-TH 0105276;%%
}
% In this paper they compute two point functions exactly for some flows.

\lref\skenderistwo{
M.~Bianchi, D.~Z.~Freedman and K.~Skenderis,
%``Holographic renormalization,''
Nucl.\ Phys.\ B {\bf 631}, 159 (2002)
[arXiv:hep-th/0112119].
%%CITATION = HEP-TH 0112119;%%
}
% In this paper they do a more general analysis.

\lref\mukhanov{
V. Mukhanov and  G. Chibisov, Zh. Eksp. Teor. Fiz. 83 (1982) 475.
}

\lref\mukhanovoriginal{
V.~F.~Mukhanov and G.~V.~Chibisov,
%``Quantum Fluctuation And 'Nonsingular' Universe. (In Russian),''
JETP Lett.\  {\bf 33}, 532 (1981)
[Pisma Zh.\ Eksp.\ Teor.\ Fiz.\  {\bf 33}, 549 (1981)].
%%CITATION = JTPLA,33,532;%%
}

\lref\starobinskyzero{
A.~A.~Starobinsky,
%``Spectrum Of Relict Gravitational Radiation And The Early State Of The  Universe,''
JETP Lett.\  {\bf 30}, 682 (1979)
[Pisma Zh.\ Eksp.\ Teor.\ Fiz.\  {\bf 30}, 719 (1979)].
%%CITATION = JTPLA,30,682;%%
}

\lref\starobinskytwo{
D.~Munshi, T.~Souradeep and A.~A.~Starobinsky,
%``Skewness of cosmic microwave background temperature fluctuations due to nonlinear gravitational instability,''
Astrophys.\ J.\  {\bf 454}, 552 (1995)
[arXiv:astro-ph/9501100].
%%CITATION = ASTRO-PH 9501100;%%
}

\lref\reviewpert{
V.~F.~Mukhanov, H.~A.~Feldman and R.~H.~Brandenberger,
%``Theory Of Cosmological Perturbations. Part 1. Classical Perturbations. Part 2. Quantum Theory Of Perturbations. Part 3. Extensions,''
Phys.\ Rept.\  {\bf 215}, 203 (1992).
%%CITATION = PRPLC,215,203;%%
}

\lref\bardeen{
J.~M.~Bardeen,
%``Gauge Invariant Cosmological Perturbations,''
Phys.\ Rev.\ D {\bf 22}, 1882 (1980).
%%CITATION = PHRVA,D22,1882;%%
}

\lref\birrel{ N. Birrel and P. Davies,
``Quantum fields in curved space'', Cambridge Univ. Press 1982.}

\lref\bs{D.~S.~Salopek and J.~R.~Bond,
%``Nonlinear Evolution Of Long Wavelength Metric Fluctuations In Inflationary Models,''
Phys.\ Rev.\ D {\bf 42}, 3936 (1990).
%%CITATION = PHRVA,D42,3936;%%
}

\lref\ks{E.~Komatsu and D.~N.~Spergel,
%``The cosmic microwave background bispectrum as a test of the physics of inflation and probe of the astrophysics of the low-redshift universe,''
arXiv:astro-ph/0012197.
%%CITATION = ASTRO-PH 0012197;%%
}

\lref\ksother{
E.~Komatsu and D.~N.~Spergel,
%``Acoustic Signatures in the Primary Microwave Background Bispectrum,''
arXiv:astro-ph/0005036;
%%CITATION = ASTRO-PH 0005036;%%
%``Acoustic Signatures in the Primary Microwave Background Bispectrum,''
arXiv:astro-ph/0005036.
%%CITATION = ASTRO-PH 0005036;%%
}

\lref\komatsucobe{
E.~Komatsu, B.~D.~Wandelt, D.~N.~Spergel, A.~J.~Banday and K.~M.~Gorski,
%``Measurement of the cosmic microwave background
bispectrum on the COBE DMR sky maps,''
Astrophys.\ J.\  {\bf 566}, 19 (2002)
[arXiv:astro-ph/0107605].
%%CITATION = ASTRO-PH 0107605;%%
}

\lref\komatsu{E.~Komatsu,
%``The Pursuit of Non-Gaussian Fluctuations in the Cosmic Microwave Background,''
arXiv:astro-ph/0206039.
%%CITATION = ASTRO-PH 0206039;%%
}

\lref\othervacua{E.~Mottola,
%``Particle Creation In De Sitter Space,''
Phys.\ Rev.\ D {\bf 31}, 754 (1985).
%%CITATION = PHRVA,D31,754;%%
B.~Allen,
%``Vacuum States In De Sitter Space,''
Phys.\ Rev.\ D {\bf 32}, 3136 (1985).
%%CITATION = PHRVA,D32,3136;%%
}

\lref\othersmall{
R.~Easther, B.~R.~Greene, W.~H.~Kinney and G.~Shiu,
%``Imprints of short distance physics on inflationary cosmology,''
arXiv:hep-th/0110226.
%%CITATION = HEP-TH 0110226;%%
N.~Kaloper, M.~Kleban, A.~Lawrence, S.~Shenker and L.~Susskind,
%``Initial conditions for inflation,''
arXiv:hep-th/0209231.
%%CITATION = HEP-TH 0209231;%%
U.~H.~Danielsson,
%``A note on inflation and transplanckian physics,''
Phys.\ Rev.\ D {\bf 66}, 023511 (2002)
[arXiv:hep-th/0203198].
%%CITATION = HEP-TH 0203198;%%
R.~Easther, B.~R.~Greene, W.~H.~Kinney and G.~Shiu,
%``A generic estimate of trans-Planckian modifications to the primordial  power spectrum in inflation,''
Phys.\ Rev.\ D {\bf 66}, 023518 (2002)
[arXiv:hep-th/0204129].
%%CITATION = HEP-TH 0204129;%%
N.~Kaloper, M.~Kleban, A.~E.~Lawrence and S.~Shenker,
%``Signatures of short distance physics in the cosmic microwave  background,''
arXiv:hep-th/0201158.
%%CITATION = HEP-TH 0201158;%%References on other vacua
}

\lref\paynecaroll{T.~Pyne and S.~M.~Carroll,
%``Higher-Order Gravitational Perturbations of the Cosmic Microwave Background,''
Phys.\ Rev.\ D {\bf 53}, 2920 (1996)
[arXiv:astro-ph/9510041].
%%CITATION = ASTRO-PH 9510041;%%
}

\lref\luo{
X.~c.~Luo and D.~N.~Schramm,
%``Testing for the Gaussian nature of cosmological density perturbations  through the three-point temperature correlation function,''
Phys.\ Rev.\ Lett.\  {\bf 71}, 1124 (1993)
[arXiv:astro-ph/9305009].
%%CITATION = ASTRO-PH 9305009;%%
}

\lref\wittentop{
E.~Witten,
%``AdS/CFT correspondence and topological field theory,''
JHEP {\bf 9812}, 012 (1998)
[arXiv:hep-th/9812012].
%%CITATION = HEP-TH 9812012;%%
}

\lref\polchinski{
J. Polchinski, ``Superstring theory', Vol 1,
 Cambridge University Press 1998 }

\lref\larsen{
F.~Larsen, J.~P.~van der Schaar and R.~G.~Leigh,
%``de Sitter holography and the cosmic microwave background,''
JHEP {\bf 0204}, 047 (2002)
[arXiv:hep-th/0202127].
%%CITATION = HEP-TH 0202127;%%
}

\lref\skenderisreview{
K.~Skenderis,
%``Lecture notes on holographic renormalization,''
arXiv:hep-th/0209067.
%%CITATION = HEP-TH 0209067;%%
}

\lref\lyth{
D. Lyth, Phys. Rev. D 31 (1985) 1792.
}

\lref\wittends{E.~Witten,
%``Quantum gravity in de Sitter space,''
arXiv:hep-th/0106109.
%%CITATION = HEP-TH 0106109;%%
}

\lref\stromingerds{A.~Strominger,
%``The dS/CFT correspondence,''
JHEP {\bf 0110}, 034 (2001)
[arXiv:hep-th/0106113].
%%CITATION = HEP-TH 0106113;%%
}

\lref\review{
O.~Aharony, S.~S.~Gubser, J.~M.~Maldacena, H.~Ooguri and Y.~Oz,
%``Large N field theories, string theory and gravity,''
Phys.\ Rept.\  {\bf 323}, 183 (2000)
[arXiv:hep-th/9905111].
%%CITATION = HEP-TH 9905111;%%
}

\lref\vijaycounterterms{
V.~Balasubramanian, J.~de Boer and D.~Minic,
%``Mass, entropy and holography in asymptotically de Sitter spaces,''
Phys.\ Rev.\ D {\bf 65}, 123508 (2002)
[arXiv:hep-th/0110108].
%%CITATION = HEP-TH 0110108;%%
}

\lref\otherds{D.~Klemm,
%``Some aspects of the de Sitter/CFT correspondence,''
Nucl.\ Phys.\ B {\bf 625}, 295 (2002)
[arXiv:hep-th/0106247].
%%CITATION = HEP-TH 0106247;%%
R.~Bousso, A.~Maloney and A.~Strominger,
%``Conformal vacua and entropy in de Sitter space,''
Phys.\ Rev.\ D {\bf 65}, 104039 (2002)
[arXiv:hep-th/0112218].
%%CITATION = HEP-TH 0112218;%%
M.~Spradlin and A.~Volovich,
%``Vacuum states and the S-matrix in dS/CFT,''
Phys.\ Rev.\ D {\bf 65}, 104037 (2002)
[arXiv:hep-th/0112223].
%%CITATION = HEP-TH 0112223;%%
V.~Balasubramanian, J.~de Boer and D.~Minic,
%``Exploring de Sitter space and holography,''
arXiv:hep-th/0207245.
%%CITATION = HEP-TH 0207245;%%
}

\lref\susskind{
L.~Dyson, J.~Lindesay and L.~Susskind,
%``Is there really a de Sitter/CFT duality,''
JHEP {\bf 0208}, 045 (2002)
[arXiv:hep-th/0202163].
%%CITATION = HEP-TH 0202163;%%
}

\lref\danielson{
U.~H.~Danielsson,
%``Inflation, holography and the choice of vacuum in de Sitter space,''
JHEP {\bf 0207}, 040 (2002)
[arXiv:hep-th/0205227].
%%CITATION = HEP-TH 0205227;%%
}

\lref\dbvv{
J.~de Boer, E.~Verlinde and H.~Verlinde,
%``On the holographic renormalization group,''
JHEP {\bf 0008}, 003 (2000)
[arXiv:hep-th/9912012].
%%CITATION = HEP-TH 9912012;%%
}

\lref\hh{
J.~B.~Hartle and S.~W.~Hawking,
%``Wave Function Of The Universe,''
Phys.\ Rev.\ D {\bf 28}, 2960 (1983).
%%CITATION = PHRVA,D28,2960;%%
}

\lref\periwal{
G.~Lifschytz and V.~Periwal,
%``Schwinger-Dyson = Wheeler-DeWitt: Gauge theory observables as bulk  operators,''
JHEP {\bf 0004}, 026 (2000)
[arXiv:hep-th/0003179].
%%CITATION = HEP-TH 0003179;%%
}

\lref\nongaussian{
D.~S.~Salopek,
%``Cold dark matter cosmology with nonGaussian fluctuations from inflation,''
Phys.\ Rev.\ D {\bf 45}, 1139 (1992).
%%CITATION = PHRVA,D45,1139;%%
F.~Bernardeau and J.~P.~Uzan,
%``Non-Gaussianity in multi-field inflation,''
arXiv:hep-ph/0207295.
%%CITATION = HEP-PH 0207295;%%
N.~Bartolo, S.~Matarrese and A.~Riotto,
%``Non-Gaussianity from inflation,''
Phys.\ Rev.\ D {\bf 65}, 103505 (2002)
[arXiv:hep-ph/0112261].
%%CITATION = HEP-PH 0112261;%%
A.~D.~Linde and V.~Mukhanov,
%``Nongaussian isocurvature perturbations from inflation,''
Phys.\ Rev.\ D {\bf 56}, 535 (1997)
[arXiv:astro-ph/9610219].
%%CITATION = ASTRO-PH 9610219;%%
D.~H.~Lyth, C.~Ungarelli and D.~Wands,
%``The primordial density perturbation in the curvaton scenario,''
arXiv:astro-ph/0208055.
%%CITATION = ASTRO-PH 0208055;%%
}

\lref\kw{
I.~R.~Klebanov and E.~Witten,
%``AdS/CFT correspondence and symmetry breaking,''
Nucl.\ Phys.\ B {\bf 556}, 89 (1999)
[arXiv:hep-th/9905104].
%%CITATION = HEP-TH 9905104;%%
}

\lref\wittendouble{
E.~Witten,
%``Multi-trace operators, boundary conditions, and AdS/CFT correspondence,''
arXiv:hep-th/0112258.
%%CITATION = HEP-TH 0112258;%%
}

%%%%%%%%%%%%%%%%%%%%%%%%%%%%%%%%%%%%%%%%%%%%%%%%%%%%%%%%%%%%%%%%%%

{ \Title{\vbox{\baselineskip12pt
\hbox{astro-ph/0210603}
%\hbox{ HUTP-97/A097}
}}
{\vbox{
{\centerline { Non-Gaussian features of primordial fluctuations }}
{\centerline { in single field inflationary models }}
  }} }
\bigskip
\centerline{ Juan Maldacena }
\bigskip
\centerline{ Institute for Advanced Study}
\centerline{Princeton, NJ 08540,USA}
\bigskip

\vskip .3in

We compute the three point correlation functions for primordial
scalar and tensor  fluctuations in single field
inflationary models. We obtain explicit expressions in the slow
roll limit where the answer is given terms of the two usual slow roll
parameters.
In a particular limit
the three point functions are determined completely by  the tilt of the
spectrum of the two point functions. We also make some remarks on
the relation of this computation to dS/CFT and  AdS/CFT. We emphasize
that (A)dS/CFT can be viewed as a statement about the wavefunction of
the universe.

% \Date{December  2000}
\vfill
\eject

%%%%%%%%%%%%%%%%%%%%%%%%%%%%%%%%%%%%%%%%%%%%%%%%%%%%%%%%%%%%%%%%%%

\newsec{Introduction and summary of results}

Single field inflationary models
predict to a good approximation a Gaussian
spectrum of primordial fluctuations.  The size of non-gaussian
corrections is  expected to be small and was estimated in
 \wise
\srednicki  \gangui  \acquaviva .

In this paper we will  compute  the correction to the Gaussian
answer to leading order in the slow roll parameters but with
the precise numerical coefficient as well as momentum dependence.
In single
field inflationary models one considers the action of a single
scalar field coupled to gravity. This action is expanded around a
spatially homogeneous solution. The leading order term in the
expansion is quadratic in the small fluctuations around the
homogeneous answer. Since a free field is a collection of harmonic
oscillators and these harmonic oscillators start their life in the
ground state one finds that the fluctuations are gaussian to
leading order. The non-Gaussian effects come from the  cubic
interaction terms in the full action. These interaction terms
arise from the non-linearities of the Einstein action as well as
from non-linearities in the potential for a scalar field.  We
compute the cubic terms in the lagrangian. These cubic terms
 lead to a change both in the ground state of the quantum
field as well as non-linearities in the evolution. These two
effects can be computed in a simple way by following the usual
rules of quantum field theory, assuming the standard choice of
vacuum for an interacting field.

 We parameterize the scalar fluctuations
in terms of $\zeta$ which is the  gauge invariant variable that
remains constant outside the horizon \bst . We schematically
denote by $\gamma$ the tensor (or gravity wave) fluctuations.
In the slow roll approximation we obtain
\eqn\summaryres{\eqalign{
 \langle \zeta_{\vec k_1}\zeta_{\vec k_2} \zeta_{\vec k_3} \rangle  =&
 { H^4 \over M_{pl}^4} { 1 \over \epsilon}
\delta^3( \sum \vec k_i) { \cal M}_1 \cr
 \langle \zeta_{\vec k_1} \zeta_{\vec k_2} h_{\vec k_3} \rangle = &
 { H^4 \over M_{pl}^4} { 1 \over \epsilon}
\delta^3( \sum \vec k_i) { \cal M}_2 \cr
 \langle \zeta_{ \vec k_1} \gamma_{\vec k_2}  \gamma_{\vec k_3} \rangle = &
 { H^4 \over M_{pl}^4}
\delta^3( \sum \vec k_i) { \cal M}_3 \cr
 \langle \gamma_{\vec k_1} \gamma_{\vec k_2}  \gamma_{\vec k_3} \rangle = &
 { H^4 \over M_{pl}^4}
\delta^3( \sum \vec k_i) { \cal M}_4
}}
 where $\epsilon$ is a
slow roll parameter and ${\cal M}_i$ are homogeneous functions of
the momenta of degree $ k^{-6}$ whose explicit form we give below.
The dependence on
 $H/M_{pl}$ is due to the fact that we are looking at the cubic term
 in the action.
% There is a very simple way to understand the form of \summaryres .
Of course, the power of $k^{-6}$ in
${\cal M}_i$ comes from approximate scale invariance.

In the limit that one of the momenta in \summaryres\ is much smaller
than the other two there is a simple argument which determines
the three point functions. This simple argument can also be used to
understand the factors of slow roll parameters in \summaryres .
The argument is the following.
Consider the limit $k_1 \ll k_{2,3}$.
Suppose the small momentum corresponds to a scalar fluctuation
$\zeta$. The fluctuation  $\zeta_{k_1}$ is
 frozen by the time the other two momenta cross the
horizon. So its only effect is to
 rescale the other two momenta  so that we
get a contribution proportional to the violation in scale invariance
of the two point function of the two fluctuations with large momenta.
 In other words the first and
third line of \summaryres\ are proportional to the tilt of the
scalar and tensor fluctuations respectively (times the product
of the corresponding two point functions).
More explicitly,
for the three $\zeta$ correlator we get\foot{ We are dropping
some  factors of $(2 \pi)^3 \delta ( \sum \vec k )$. These
are more explicitly written later.}
\eqn\resthree{\eqalign{
\langle \zeta_{\vec k_1} \zeta_{\vec k_2} \zeta_{\vec k_3} \rangle \sim &
 -  \langle \zeta_{\vec k_1}
  \zeta_{-\vec k_1} \rangle k {d \over d k }
\langle  \zeta_{\vec k_2} \zeta_{ \vec k_3} \rangle  = \cr
 =&- n_s \langle  \zeta_{\vec k_2} \zeta_{ \vec k_3} \rangle
  \langle \zeta_{\vec k_1}
  \zeta_{-\vec k_1} \rangle ~,~~~~~~~~~~~~ k_1 \ll k_{2}, k_3
}}
where $n_s$ is the tilt of the scalar spectrum defined by
$\langle \zeta \zeta \rangle \sim k^{-3 + n_s}$ so that $n_s$ is
the deviation from scale invariance.

In order to understand the behavior when $k_i$ are all of the same
order of magnitude we need to do the computation by expanding the action.
Then the answer is a more complicated function of $k_i$ but the size
of the correlation function does not numerically change much.
 The other two correlation functions in \summaryres\ can also
 be understood in the limit that one of the $k_i$ is very small
 through a similar simple argument which we give in section 4.

Another way of presenting the argument is as follows.
Since the wavefunction of gravity in a space that is approximately
de-Sitter is  supposed to have the properties of a conformal field theory
\wittends \stromingerds ,
the three point functions that we computed above can be related to
correlation functions of the stress tensor in the hypothetical
dual CFT. In the limit that
one of the $k_i$ is much smaller than the other two the form of
the three point function is determined in terms of the two point function
by the following argument.
If one of the $k_i$ is very small we can approximate it by zero,
so that
 the corresponding insertion of the trace of the
stress tensor represents the effects of an infinitesimal
rescaling  of coordinates. So  this three point function is determined
by how the two point function behaves as we rescale the  coordinates.
This is why the three point function $\langle \zeta^3 \rangle$
is equal to the tilt of the spectrum of the two point
function in the regime $k_1 \ll k_{2,3}$.

Komatsu and Spergel have performed an analysis of the detectability
of
non-gaussian features of the temperature fluctuations \ks \ksother .
Their analysis was made for an expected signal which had a
slightly different  $k$ dependence from the one in ${\cal M}_1$
above. This probably would not change their answer too much.
Ignoring this point, one would conclude from their analysis that
this level of non-gaussianity is not detectable from CMB measurements
alone. A more explicit discussion is given below.
In some models with more than one field non-gaussianity can be
large \nongaussian .

Finally we point out that these computations can also be used in
investigations of AdS/CFT and dS/CFT. These dualities can be viewed
as a statement about the wavefunction of the universe. We relate
explicitly the computation of stress tensor correlators in the
dS and AdS case. They are related  by a simple analytic continuation.
We also clarify the relation between  stress tensor correlators
and the spectrum of fluctuations of metric perturbations.

This paper is organized as follows. In section two we
review the standard results that follow from the quadratic
approximation and give the gaussian answer. In section three
we expand the action to third order. In section four we compute the
three point functions.
In section five  we make some remarks on the relationship
of these computations  to the dS/CFT and AdS/CFT correspondences.

\newsec{ Review of the quadratic computation}

The computation of primordial fluctuations that arise
in inflationary models was first discussed in
\mukhanovoriginal
\starobinsky \hawking \guth
\bst \mukhanov\ and was nicely reviewed in \reviewpert .

The starting point is the Lagrangian of gravity and a scalar field which
has the general form
\eqn\lagrang{
S = { 1 \over 2}  \int \sqrt{g} [ R - (\nabla \phi)^2 - 2 V(\phi)]
}
up to field redefinitions. We have set $M_{pl}^{-2} \equiv
8 \pi G_N =1$\foot{ Note that this  definition of $M_{pl}$
is different from the definition
that some other authors use
(including Planck).}, the dependence
on $G_N$ is easily reintroduced.

The homogeneous solution  has the form
\eqn\homo{
ds^2 = - dt^2 + e^{ 2 \rho(t)} dx_i dx_i  = e^{ 2 \rho} ( - d\eta^2 +
dx_i dx_i )
}
where $\eta$ is conformal time.
The scalar field is a function of time only.
 $\rho$ and $\phi$ obey
the equations
\eqn\homoeq{\eqalign{
3 \dot \rho^2  =& { 1 \over 2} \dot \phi^2 + V(\phi)
\cr
\ddot{\rho} = & - { 1 \over 2} \dot \phi^2
\cr
0= & \ddot{ \phi} + 3 \dot \rho \dot \phi + V'(\phi)
}}
The Hubble parameter is $H \equiv \dot \rho$. The third equation follows
from the first two. We will make frequent use of these equations.

If the slow roll parameters are small we will have a period of
accelerated expansion.
The slow roll parameters are defined as
\eqn\slowroll{\eqalign{
\epsilon & \equiv { 1 \over 2 }
\left( { M_{pl} V' \over V} \right)^2 \sim
 {1\over 2} { \dot \phi^2 \over \dot \rho^2 } {1 \over M_{pl}^2 }
\cr
\eta & \equiv
{M_{pl}^2  V'' \over V} \sim
- { \ddot \phi \over \dot \rho \dot \phi } + { 1 \over 2} { \dot \phi^2
\over \dot \rho^2 } { 1\over M_{pl}^2 }
}}
where the approximate relations hold when the slow roll parameters
are small.

We now consider small fluctuations around the solution \homoeq .
We expect to have three physical propagating degrees of freedom,
two from gravity and one from the scalar field. The scalar field
mixes with other components of the metric which are also scalars
under $SO(2)$ (the little group that leaves $\vec k $ fixed).
 There are four scalar modes of the metric which
are  $\delta g_{00}$, $\delta g_{ii}$,
$\delta g_{0i} \sim \partial_i B $ and
$\delta g_{ij} \sim \partial_i \partial_j H$ where $B$ and $H$ are
arbitrary functions. Together with a small fluctuation,
$\delta \phi$,  in the scalar field
these total  five scalar modes. The action \lagrang\ has
gauge invariances coming from reparametrization invariance. These
can be linearized for small fluctuations. The scalar modes are
acted upon by
two gauge invariances, time reparametrizations and
spatial reparametrizations of the form $x^i \to x^i + \epsilon^i(t,x)$
with $\epsilon^i = \partial_i \epsilon$. Other coordinate
transformations act on the vector modes\foot{There are
no propagating vector modes for this Lagrangian \lagrang .
 They are removed by
gauge invariance and the constraints. Vector modes are present when
more fields are included.}. Gauge invariance
removes two of the five functions. The constraints in the action
remove two others  so that we are left with one degree of freedom.

In order to proceed it is convenient to work in the ADM formalism.
We write the metric as
\eqn\metric{
ds^2 = - N^2 dt^2 + h_{ij} (dx^i + N^i dt)(dx^j + N^j dt)
}
and the action \lagrang\ becomes
\eqn\adm{
S = { 1 \over 2} \int \sqrt{h} \left[ N R^{(3)} - 2 N V +
N^{-1} ( E_{ij} E^{ij} - E^2) + N^{-1}
( \dot \phi - N^i \partial_i \phi)^2  - N h^{ij} \partial_i \phi
\partial_j \phi
\right]
}
Where
\eqn\defofe{\eqalign{
E_{ij} & = { 1 \over 2} ( \dot h_{ij} - \nabla_i N_j - \nabla_j N_i)
\cr
E & = E_i^i
}}
Note that the extrinsic curvature is $K_{ij} = N^{-1} E_{ij}$.
In the computations we do below it is often convenient  to separate
the traceless and the trace part of $E_{ij}$.

In the ADM formulation  spatial coordinate reparametrizations are
an explicit symmetry while time reparametrizations are not
so obviously a symmetry.
The ADM formalism is designed so that one can think of
$h_{ij}$ and $\phi$ as the dynamical variables and $N$ and $N^i$
as Lagrange multipliers. We will choose a gauge for $h_{ij}$ and
$\phi$ that will fix time and spatial reparametrizations.
A  convenient gauge is
\eqn\gauge{
\delta \phi= 0 ~, ~~~~~~  h_{ij} = e^{ 2 \rho }
[ (1 + 2 \zeta) \delta_{ij} + \gamma_{ij}  ] ~,~~~~~~~ \partial_i \gamma_{ij} =0
~,~~~~~~~ \gamma_{ii} =0
}
where $\zeta$ and $\gamma$ are first order quantities.
$\zeta$ and $\gamma$ are the  physical degrees of freedom. $\zeta$
parameterizes the scalar fluctuations and $\gamma$ the tensor
fluctuations.
The gauge \gauge\ fixes the gauge completely at nonzero momentum.
In order to find the action for these degrees of freedom we just
solve for $N$ and $N^i$ through their equations of motion and
plug the result back in the action. This procedure gives
the correct answer since $N$ and $N^i$ are Lagrange multipliers.
The gauge \gauge\ is very similar to Coulomb gauge in electrodynamics
where we set $\partial_i A_i =0$,  solve for $A_0$ through its
equation of motion and plug this back  in the action\foot{
As in electrodynamics in Coulomb gauge we will often find expressions
which are not local in the spatial  directions. In the linearized theory
it is possible to define local gauge invariant observables where these
non-local terms disappear. }.

The equation of motion for $N^i$ and $N$ are the
 the momentum and hamiltonian constraints
\eqn\constr{\eqalign{
 ~ & \nabla_i [ N^{-1} ( E^i_j - \delta^i_j E) ] =0
\cr
 ~ & R^{(3)} - 2 V - N^{-2} ( E_{ij} E^{ij} - E^2 ) - N^{-2} \dot \phi^2 =0
}}
where we have used that $\delta \phi =0$ from \gauge .
We can solve these equations to first order by setting
$N^i = \partial_i \psi + N_T^i $ where $\partial_i N^i _T =0$ and
$N = 1+ N_1$.
We find
\eqn\first{
N_1 = { \dot \zeta \over \dot \rho}~,~~~~~~~~~~~~N^i_T =0
  ~,~~~~~~~ \psi =
 - e^{-2 \rho}
{ \zeta \over \dot \rho} + \chi ~,~~~~~~~~ \partial^2 \chi=
 { \dot \phi^2 \over
2 \dot \rho^2 } \dot \zeta
}

In order to find the quadratic action for $\zeta$ we can replace
\first\ in the action and expand the action to second order. For
this purpose it is not necessary to compute $N$ or $N^i$ to second
order. The reason is that the second order term in $N$ will be
multiplying the hamiltonian constraint, ${\partial L \over
\partial N}$ evaluated to zeroth order which vanishes since the
zeroth order solution obeys the equations of motion. There is a
similar argument for $N^i$. Direct replacement in the action
gives, up to second order, \eqn\expr{\eqalign{ S =& { 1 \over 2}
\int e^{ \rho + \zeta}(1 + { \dot \zeta \over \dot \rho}) [ - 4
\partial^2 \zeta - 2 (\partial \zeta)^2 - 2 V e^{ 2 \rho + 2 \zeta} ]   + \cr
  ~~& ~~ + e^{ 3 \rho + 3 \zeta}
{ 1 \over ( 1 + { \dot \zeta \over \dot \rho})} [ - 6 ( \dot \rho
+ \dot \zeta)^2
  + \dot \phi^2
]
}}
where we have neglected a total derivative which is linear in $\psi$.
After integrating by parts some of the terms and using the
background equations of motion \homoeq\  we find the
final expression to second order\foot{
In order to compare this to the expression in \reviewpert\ set
$ v = - z \zeta $ in (10.73) of \reviewpert .}
\eqn\quadraticscalar{
S = { 1 \over 2} \int  dt d^3x   { \dot \phi^2 \over \dot \rho^2}
[ e^{ 3 \rho} \dot \zeta^2 - e^\rho (\partial \zeta)^2 ]
}
No slow roll approximation was made in deriving \expr .
Note that naively the action \expr\ contains terms of the order
$\dot \zeta^2$, while the final expression contains only terms of the
form $\epsilon \dot \zeta^2$, so that the action is suppressed by a
slow roll parameter. The reason is that the $\zeta$ fluctuation
would be a pure gauge mode in de-Sitter space and it gets a
non-trivial action only to the extent that the slow roll parameter
is non-zero. So the leading order terms in slow roll in  \expr\ cancel
leaving only the terms in \quadraticscalar . A simple argument for the
dependence of \quadraticscalar\ on the slow roll parameters is given
below.

Since \quadraticscalar\ is describing a free field we just
have a collection of harmonic oscillators.
More precisely we expand
\eqn\fourexpa{
\zeta(t,x) = \int { d^3 k \over ( 2 \pi)^3 } \zeta_k(t)
e^{i \vec k \vec x}
}

Each $\zeta_k(t)$ is a harmonic oscillator with time
dependent  mass and spring
constants. The quantization is straightforward \birrel .
We pick two
independent
classical solutions $\zeta_k^{cl}(t) $ and $\zeta_{k}^{cl *}(t)$
of the equations of motion of \quadraticscalar\
\eqn\classeq{
{ \delta L \over \delta \zeta} =  - { d \left( e^{ 3 \rho} { \dot \phi^2 \over \dot \rho^2} \dot \zeta_k
\right) \over dt }  -  { \dot \phi^2 \over \dot \rho^2}
 e^ \rho k^2 \zeta_k = 0
}
Then we write
\eqn\zetaexpr{
\zeta_{\vec k}(t) =  \zeta_k^{cl}(t) a^\dagger_{\vec k}  +
  \zeta_k^{cl*}(t) a_{-\vec k}
}
where $a$ and $a^\dagger$ are some operators.
Demanding that $a^\dagger$ and $a$ obey the standard creation
and annihilation commutation relations we get a normalization
condition for $\zeta^{cl}_k$. Different choices of solutions are
different choices of vacua for the scalar field.
The comoving wavelength of each mode
 $\lambda_c \sim 1/k$ stays constant but
the physical wavelength changes in time.
For early times the ratio of the physical wavelength to the
Hubble scale is very small and the mode feels it is in almost  flat
space. We can then use the WKB approximation to
solve \classeq\ and choose the usual vacuum in Minkowski space.
When the physical wavelength is much longer than the Hubble
scale
\eqn\separcond{
 \lambda_{phys} H =    { \dot \rho   e^{\rho} \over k } \gg 1
}
the solutions of \classeq\ go rapidly to a constant.

A useful example to keep in mind is that of a massless scalar
field $f$ in de-Sitter
space.
In that case the action is $S = { 1\over 2} \int H^{-2} \eta^{-2}
[ (\partial_\eta f)^2 - (
\partial f)^2 ] $ and the normalized classical solution, analogous
to $\zeta^{cl}_k$, corresponding to the standard Bunch Davies vacuum is
\birrel\
\eqn\classsolds{
f^{cl}_k = { H \over \sqrt{ 2 k^3} } ( 1 - i { k \eta })
e^{ i {k \eta }}
}
where we are using conformal time which runs from $(- \infty, 0)$.
Very late times correspond to small $|\eta|$ and we clearly see from
\classsolds\ that $f^{cl}$ goes to a constant. Any solution,
including \classsolds ,  approaches
a  constant at late times   as $ \eta^2\sim e^{- 2\rho} $,
which is exponentially
fast is physical time.
In de-Sitter space we can  easily compute the two
point function for this scalar field and obtain\foot{
In coordinate space the result for late times
 is $\langle f(x,t) f(x' ,t) \rangle
sim - { H^2 \over ( 2 \pi)^2} \log( |x-x'|/L) $ where is an
IR cutoff which is unimportant when we compute differences in $f$ as
we do in actual experiments. }
\eqn\twods{\eqalign{
\langle f_{\vec k}(\eta)  f_{\vec k'}(\eta) \rangle = & ( 2 \pi)^3
\delta^3( \vec k + \vec k') | f^{cl}_{k}(\eta)|^2 =
 ( 2 \pi)^3
\delta^3( \vec k + \vec k') { H^2 \over 2 k^3} (1 + k^2 \eta^2  )
\cr
  \sim &( 2 \pi)^3
\delta^3( \vec k + \vec k') { H^2 \over 2 k^3}  ~~~~~~~
  { \rm for}~~~ k \eta \ll 1
  }}

We now go  back to the inflationary computation.
If one knew the classical solution to the equation \classeq\ the
result for the correlation function of $\zeta$ can be simply computed as
\eqn\zetacorrexact{
\langle \zeta_{\vec k} (t) \zeta_{\vec k'}(t) \rangle =
( 2 \pi)^3
\delta^3( \vec k + \vec k') | \zeta^{cl}_{ k}(t)|^2
}

If the slow roll parameters are small when the comoving scale $\vec k$
crosses the horizon then it is possible to estimate
the late time behavior of
 \zetacorrexact\ by the corresponding result in de-Sitter
space \twods\ with a Hubble constant that is the Hubble constant
at the moment of horizon crossing. The reason is that at late
times $\zeta$ is constant while at early times the field is in the
vacuum and its wavefunction is accurately given by the WKB
approximation. Since the action \quadraticscalar\ also contains a
factor of $\dot \phi/\dot \rho$ we also have to set its value to
the value at horizon crossing, this factor only appears in
normalizing the classical solution. In other words, near horizon
crossing we set $f = { \dot \phi \over \dot \rho} \zeta$ where $f
$ is a canonically normalized
 field in de-Sitter space.
This produces the well known result
\eqn\zetaapprox{
\langle \zeta_{\vec k} (t) \zeta_{\vec k'}(t) \rangle \sim
( 2 \pi)^3
\delta^3( \vec k + \vec k') { 1 \over 2 k^3} { \dot \rho^2_* \over
M_{pl}^2  }{  \dot \rho^2_* \over \dot \phi^2_* }
}
where the star means that it is evaluated at the time of horizon
crossing, i.e. at time $t_*$
such that
\eqn\stardef{
\dot \rho(t_*) e^{\rho(t_*)}  \sim  k ~.
}
The dependence of \zetaapprox\ on $t_*$ leads to additional momentum
dependence. It is conventional to parameterize this dependence
by saying that the total correlation function has the form
$k^{-3 +n_s}$ where
\eqn\defns{
n_s = k {d \over d k }
\log ( { \dot \rho_*^4 \over \dot \phi_*^2 })  \sim
 { 1 \over \dot \rho_*} { d \over d t_*}
 \log ( { \dot \rho_*^4 \over \dot \phi_*^2 })
 = -2( { \ddot \phi_* \over \dot \rho_* \dot \phi_*}
 + { \dot \phi_* \over \dot \rho_* } ) = 2( \eta - 3 \epsilon)
}

As it has been often discussed, after horizon crossing the mode
becomes classical, in the sense that the commutator
 $[\dot \zeta , \zeta] \to 0$
exponentially fast. So for measurements which only involve $\zeta$
or $\dot \zeta$  we can treat the mode as a classical variable.

After the end of inflation the field $\phi$ ceases to determine
the dynamics of the universe and we eventually go over
to the usual  hot big bang phase.
It is possible to prove \bst \reviewpert\
that $\zeta$ remains constant outside the horizon as long as
no entropy perturbations are generated and a certain condition
on the off-diagonal components of the spatial stress tensor is
obeyed\foot{The condition is $\partial_i \partial_j ( \delta T_{ij} -
{ 1 \over 3} \delta_{ij} T_{ll} ) =0 $ .}.
These conditions are obeyed if
 the universe is described by a single fluid or by a single scalar
 field.
We should mention that for a general fluid the variable $\zeta$
can be defined in terms of the three metric as above \gauge\
in the comoving gauge where $T^0_i=0$\foot{
For readers who are familiar with Bardeen's classic paper \bardeen ,
we should mention that the gauge invariant definition of
$\zeta$ is  $\zeta = h + ( { \cal H}^{-1} h' - A) { \cal H}^2/
({\cal H}^2 - \dot { \cal H }) $ where ${\cal H} = \rho'$ and primes
indicate derivatives with respect to conformal time and $h = H_L + H_T/3$ with $A,~ H_L,~ H_T$ defined in \bardeen . In circumstances
where $\zeta$ is conserved then it also reduces to the definition
in terms of Bardeen potentials in \bst , \reviewpert\  (actually
$\zeta_{here} = - \zeta_{there}$).
The gauge choice that makes $h = \zeta$ is $T^0_i =0$ or,
using the equations of motion,
$ \dot h = \dot \rho A $.}. In the case of a scalar field this
implies that $\delta \phi =0$.
This gauge is convenient conceptually
since the variable $\zeta$  is directly a function appearing in the
metric. We see that the variable $\zeta$ tells us how much the spatial
directions have expanded in the comoving gauge, so
that to linear order $\zeta$ determines the curvature of the
spatial slices $R^{(3)} = 4 k^2 \zeta $ \lyth .
This variable $\zeta$ is very useful in order to continue through
the end of inflation since it is defined throughout the evolution and
it is constant outside the horizon. An intuitive way to understand
why $\zeta$ is constant is to note that the conditions stated above
imply that two observers separated by some distance see the
universe undergoing precisely the same history. Outside the horizon
(where we can set $k=0$ in all equations) $\zeta$ is
just a rescaling of coordinates and this rescaling is a symmetry
of the equations.

Other gauges can be more convenient in order to do computations
in the slow roll approximation. A gauge that is particularly
convenient is
\eqn\gaugev{
\delta \phi \equiv \varphi(t,x) ~,~~~~~~~~~
h_{ij} = e^{ 2 \rho} ( \delta_{ij} + \gamma_{ij} ) ~,~~~~~~
\partial_i \gamma_{ij} =0 ~,~~~~~~\gamma_{ii} =0
}
where we have denoted the small fluctuation of the scalar field by
$\varphi$. In order to avoid confusion, from now on $\phi$ will denote
the background value of the scalar field and $\varphi$ will be its
deviation from the background value.
We expect that in this gauge the action will be approximately the
action of a massless scalar field $\varphi$ to leading order in
slow roll.
Indeed, we can check that the first order expressions for $N$ and
$N^i$ are
\eqn\expres{
N_{1 \, \varphi} =  { \dot \phi \over 2 \dot \rho} \varphi ~,~~~~~~
   ~  N^i_{\varphi}  = \partial_i \chi ~,~~~~~~~~~~
   \partial^2 \chi = { \dot \phi^2 \over
2 \dot \rho^2 }  { d \over d t}\left( -  { \dot \rho \over
\dot \phi} \varphi \right)
}
where the $\varphi$ subindex  reminds us that
$N_{1 \, \varphi} ,~ N^i_{\varphi}$  are computed in the gauge \gaugev .
We see that these expressions are subleading  in slow roll compared
to $\varphi$.
So in order to compute the quadratic action to lowest order in
slow roll it is enough to consider just the $(\nabla \varphi)^2$ term
in the action \lagrang\ since $V''$ is also of higher order in slow roll.
This is just the action of a massless scalar field in the
zeroth order background. We can compute the fluctuations in
$\varphi$ in the slow roll approximation
and we find  a  result similar to that of a scalar field
in de-Sitter space \twods\ where the Hubble scale is evaluated at
horizon crossing.
 After horizon crossing we can evaluate the
gauge invariant
quantity $\zeta$. This is most easily done by changing the gauge
to the gauge where $\varphi=0$.
 This can be achieved by a time reparametrization
of the form
$\tilde t = t + T$ with
\eqn\firstt{
T = - { \varphi \over { \dot \phi} }
}
where $t$ is the time in the gauge \gauge\ and $\tilde t$ is the
time in \gaugev .
After the gauge transformation \firstt , we find that the metric
in \gaugev\ becomes of the form in \gauge\ with
\eqn\formofzeta{
\zeta = \dot \rho T = - { \dot \rho \over \dot \phi} \varphi
}
Incidentally, this implies that $\chi$ in \expres\ is the same as
$\chi$ in \first .
So the correlation function for $\zeta$ can be computed as
the correlation function for $\varphi$ times the factor in
\formofzeta . In order to get a result as accurate as possible
we should perform the gauge transformation \formofzeta ,
just after crossing the horizon so that the factor in \formofzeta ,
is evaluated at horizon crossing leading finally to \zetaapprox .
In principle we could compute
 $\zeta$  from  $\varphi$ at any time. If we were to
choose to do it a long time after horizon crossing
we would need to take into account that $\varphi$ changes outside
the horizon. This would require  evaluating
 the action \lagrang\ to higher order
in the slow roll parameters.
Of course, the  dependence for
$\varphi$ outside the horizon is such that it
 precisely cancels the
time dependence of the  factor in
\formofzeta\ so that $\zeta $ is constant.

In summary, the computation is technically simplest if
we start with the  gauge \gaugev\ and we compute the two point
function of $\varphi$ after
horizon exit and at that time  compute the $\zeta$ variable which then
remains constant.  On the other hand
the computation in the  gauge  \gauge\ is
conceptually simpler since the whole computation always involves
the variable of interest
which is $\zeta$.
In other words, the  gauge \gaugev\ is more useful before and during
horizon
crossing while the  gauge \gauge\ is more useful after horizon
crossing.

These  last few   paragraphs
are basically
simple argument presented in \hawking .  The computation of
fluctuations of $\varphi$ in  de-Sitter produces fluctuations of
the order $\varphi = {H \over 2 \pi} $ and then this leads
to a delay in the evolution  by $\delta t= - \varphi/\dot \rho $
(see  \firstt )
which in turn gives an additional expansion of the universe
by a factor $\zeta =  \dot \rho\delta t
 = - { \dot \rho \over \dot
\phi } \varphi $.
This additional expansion is evaluated at horizon crossing in order
to minimize the error in the approximation.

We now summarize the discussion of gravitational waves
 \starobinskyzero .
Inserting \gauge\  in the
action  and focusing on terms quadratic in
$\gamma$ gives
\eqn\gravwv{
S= { 1\over 8 } \int
[e^{ 3 \rho} \dot \gamma_{ij} \dot
\gamma_{ij} - e^\rho  \partial_l  \gamma_{ij} \partial_l \gamma_{ij} ]
}
As usual we can expand $\gamma$ in plane waves with definite
polarization tensors
\eqn\expans{
\gamma_{ij} = \int {d^3k \over ( 2 \pi)^3} \sum_{s=\pm} \epsilon_{ij}^s(k)
\gamma^s_{\vec k}(t) e^{ i \vec k \vec x }
}
where $\epsilon_{ii} = k^i \epsilon_{ij} =0$ and $\epsilon^s_{ij}(k)
\epsilon^{s'}_{ij}(k)  = 2 \delta_{s s'}$.
So we see that for each polarization mode we have essentially the
equation of motion of a massless scalar field.
As in our previous discussion,
 the solutions become constant after crossing the
horizon.  Computing the correlator just after horizon
crossing we get
\eqn\resgr{
\langle \gamma^s_{\vec k} \gamma^{s'}_{\vec k'}   \rangle
=  (2 \pi)^3 \delta^3( \vec k + \vec k') { 1 \over
2 k^3} { 2 \dot \rho^2_* \over M_{pl}^2 }\delta_{ss'}
}
where we reinstated the $M_{pl}$ dependence.
We can similarly define the tilt of the gravitational wave
spectrum by saying that the correlation function scales as
$k^{-3 + n_t}$
where $n_t$ is given by
\eqn\defntres{
n_t = k { d \over d k} \log \dot \rho_*^2  = - { \dot \phi^2_*
\over \dot \rho^2_* } = - 2 \epsilon
}

\newsec{Cubic terms in the Lagrangian }

In this section we
 compute the cubic terms in the lagrangian in
two different gauges. We do this as a check of our computations.
The first gauge is similar to \gauge , which
 is conceptually simpler since
one works from the very beginning with the $\zeta$ variable
in terms of which one wants to compute the answer.
We need to fix the gauge to second order in small fluctuations.
We achieve this by  setting to zero the fluctuations in
 $\phi$ and
we writing  the 3-metric as
\eqn\firstg{\eqalign{
\delta \phi & =0
\cr
h_{ij} &= e^{ 2 \rho + 2 \zeta} \hat h_{ij} ~,~~~~~
\det  \hat h =1, ~~~~~~
 \hat h_{ij} = ( \delta_{ij} + \gamma_{ij} + { 1 \over
2} \gamma_{il} \gamma_{l j} + \cdots )
}}
where $\gamma_{ii} = \partial_i \gamma_{ij} =0$ to second order.
 The term proportional to $\gamma^2$ was introduced
 with the purpose of simplifying some formulas\foot{
We can define the gauge condition as
$\partial_i ( \log \hat h)_{ij} =0$. }.
Note that it is necessary to define $h_{ij}$ only to second order since
any third order term in $h_{ij}$ will not contribute to the action.

The second gauge that we  choose is
\eqn\secondg{\eqalign{
\phi & = \phi(t) + \varphi(t,x)
\cr
h_{ij} & = e^{2 \rho} \hat h_{ij} ~,~~~~\det \hat h =1 ~,~~~~~~~
 \hat h_{ij} =
  ( \delta_{ij} + \tilde \gamma_{ij} + { 1 \over
2} \tilde \gamma_{il} \tilde \gamma_{l j} + \cdots )
}}
again with $\tilde \gamma_{ii} = \partial_i \tilde \gamma_{ij} =0$.
In  appendix A we work out in detail the change of gauge which gives
 the
relation between the $\zeta,\gamma_{ij}$ variables and the
$\varphi, \tilde \gamma_{ij}$ variables. Here we summarize that discussion.
 We denote by
$\tilde t$ the time in the gauge \secondg\ and by $t$ the
time in the gauge \firstg . $t$ and $\tilde t$ are
related by a time reparametrization of the form
 $\tilde t = t + T(t,x)$. The function $T$ should be such that
 from $\delta \phi \not =0$ in \secondg\ we get
$\delta \phi =0$
in \firstg . This determines
\eqn\valueofT{
T = - { \varphi \over \dot \phi} - { 1 \over 2} {
\ddot \phi  \varphi^2 \over \dot \phi^3 } + { \dot \varphi \varphi
\over \dot \phi^2 }
}
Starting from \secondg\
this time reparametrization  produces a new metric.
A spatial reparametrization then    carries it to the form in \firstg\
to second order.
After all these steps we find the
following relation between the $\varphi$ and $\zeta$ variables (for
$\tilde \gamma =0$)
\eqn\changeofgauge{\eqalign{
\zeta = & \dot \rho T + { 1 \over 2} \ddot \rho T^2 - { 1 \over 4}
\partial_i T \partial_j T e^{ - 2 \rho} +
{ 1 \over 2} \partial_i \chi \partial_i T + \cr
 & + { 1\over 4} e^{ - 2 \rho} \partial^{-2} \partial_i
\partial_j \left( \partial_i T \partial_j T \right)
- { 1 \over 2}\partial^{-2} \partial_i
\partial_j \left( \partial_i \chi  \partial_j T \right)
}}
where $T$ is defined in \valueofT\ and $\chi$ is defined in \expres .
This expression will be useful for comparing results computed in
different gauges.
The change of variables relating $\tilde \gamma_{ij}$ and $\gamma_{ij}$
can be found in  appendix A.

\subsec{ Evolution outside the horizon to all orders
%\footnote{This subsection was added on
%May 2005, to replace a previous argument that was not correct (as pointed out to me by
%S. Weinberg).}
}

It is possible to show that $\zeta $ and $\gamma$ (defined in \firstg)
are constant outside the
horizon. For this purpose it is enough to expand the action to first order in
derivatives of the fields but to all orders in powers of the fields.
We will assume that $N = 1 + \delta N$ where $\delta N$ has an expansion in derivatives
that starts with a first order term. Similarly we will assume that $N^i$ is of zeroth
order in derivatives, that that $\nabla_j N_i$ are of first order in derivatives. These
assumptions are consistent with the structure of the hamiltonian and momentum
constraints \constr .
We can then expand the Hamiltonian constraint to first order in derivatives and solve
for $\delta N$  to first order in derivatives and all orders in powers of the fields\foot{
Note that if we  expand this to linear order in the fields
it only agrees with our linearized expressions \first\ to first order in derivatives.}
\eqn\solved{
 2 V \delta N =
  2 \dot \rho ( 3 \dot \zeta - \nabla_i N^i )
}
We can now evaluate the action. On a solution of the Hamiltonian constraint the action
reads
\eqn\acsolh{\eqalign{
S = &\int \sqrt{h} N (R^{3} - 2 V) = \int \sqrt{h} ( - 2 V - 2 V \delta N)  =
\cr =&
\int e^{ 3 \rho + 3 \zeta} ( - 6 \dot \rho^2 +   \dot \phi^2 - 6 \dot \zeta \dot \rho )
= - 2 \int d_t\left( e^{ 3 \rho + 3 \zeta } \dot \rho \right)
}}
where we neglected the term involving $R^{(3)}$ because it
is of second order in derivatives,
we used \solved ,  we integrated by parts the term involving $\nabla_i N^i$ and
we used \homoeq . Therefore \acsolh\ is a total derivative and can be ignored.

In conclusion, $\zeta $ and $\gamma$ are constant outside the horizon.
 The reason is that outside the
horizon we can neglect all spatial derivatives. Since we also showed that
the expansion in powers of time derivatives starts at second order we conclude
that constant $\zeta$ and $\gamma$ are solutions of the equations of
motion  to all orders in powers of $\zeta, ~\gamma$ outside the horizon.
This fact was derived in a different way in \bs
\foot{  \bs\ call $\alpha$ our $\zeta$. }.
Of course, the intuitive explanation of this fact is clear, after
exiting the horizon all regions evolve in the same fashion, the only
difference between them is how much one has expanded relative to the
other. This fact makes it clear that the definition of $\zeta$ in
\firstg\ \bs\ is the correct non-linear generalization  of  the variable
introduced in \bst .

\subsec{Cubic term for three scalars}

We now expand the action up to cubic order in  $\zeta$.
  It turns out that it is only  necessary to know
$N$ or $N^i$ up to first order. The third order terms in $N$ ,
$N^i$ would be multiplying the constraints evaluated at zeroth
order. The second order terms in $N$, $N^i$ multiply the
constraints evaluated to first order, which vanish due to the
first order expressions for $N$ and $N^i$.
Up to total derivatives
in time and space we find \eqn\cubiczetaraw{\eqalign{ S = &\int
e^{ \rho + \zeta } (1 + {\dot \zeta \over \dot \rho} ) [ - 2
\partial^2 \zeta - (\partial \zeta)^2] + e^{ 3 \rho + 3 \zeta}  {
1 \over 2} {\dot \phi^2 \over \dot \rho^2} \dot \zeta^2 ( 1 - {
\dot \zeta \over \dot \rho}) + \cr
 ~& +  e^{ 3 \rho + 3 \zeta } \left[ { 1 \over 2}
 ( \partial_i \partial_j \psi \partial_i \partial_j \psi
-(\partial^2 \psi )^2)  ( 1 - {
\dot \zeta \over \dot \rho}) - 2 \partial_i \psi \partial_i \zeta
\partial^2 \psi \right]
}}
where we expand the exponentials so that we keep only terms of up to third
order in $\zeta$, and  $\psi$ is defined in \first\ and is of first order in $\zeta$.

Something that is not obvious from \cubiczetaraw\ is that the
effective cubic interaction term is of second  order in slow roll.
Schematically  it
 is of the form $ \epsilon ^2  \dot \zeta^2 \zeta $, while
the action \cubiczetaraw\ appears to be of order $\epsilon^0 $.
Factors of $\epsilon $ are very important since they will
determine whether this effect is measurable or not \ks .

An easy way to see that the effective action is of order $\epsilon^2$
is to compute the cubic term in the action
 in the  gauge \secondg ,
which leads to
\eqn\actthr{\eqalign{
S_3 = &\int  e^{3 \rho} \left( - { \dot \phi \over 4
\dot \rho} \varphi \dot \varphi^2 -
e^{ - 2 \rho} { \dot \phi \over 4 \dot
\rho } \varphi ( \partial \varphi)^2
- \dot \varphi \partial_i \chi \partial_i \varphi
\right. +
\cr
&+ { 3 \dot \phi^3 \over 8 \dot \rho}
\varphi^3 -
{ \dot \phi^5 \over 16 \dot \rho^3 } \varphi^3 - { \dot \phi V''
\over 4 \dot \rho} \varphi^3 -
{ V''' \over 6} \varphi^3 + { \dot \phi^3 \over
4 \dot \rho^2 } \varphi^2 \dot \varphi
+ { \dot \phi^2 \over 4 \dot \rho}  \varphi^2 \partial^2 \chi
\cr
+ & \left. { \dot \phi \over 4 \dot \rho} (
- \varphi \partial_i \partial_j \chi \partial_i \partial_j \chi +
\varphi \partial^2 \chi \partial^2 \chi ) \right)
}}
Only the terms in the first line of \actthr\
contribute to leading order in
the slow roll approximation. The term proportional to $V'''$ was
considered in \srednicki\ but we see that it is subleading in the
slow roll approximation.
By using the first order relation between $\zeta$ and $\varphi$
\formofzeta\
we see that the  first line of \actthr\ leads to an effective action
which is of order $\epsilon^2$ in the $\zeta$ variables.
On the other hand, in the action \actthr , it is not obvious that
there is any variable which stays constant outside the horizon.
Indeed there are $\varphi^3$ couplings that typically lead to
evolution of the perturbations outside the horizon \srednicki.

Since one property (the constancy of $\zeta$) is clear in one gauge
while the other (the fact that the interaction is of order $\epsilon^2$)
is more clear in the other the reader might have some doubts about
one or both statements. In order to
dissipate all doubts about these statements
we start from the cubic term in $\zeta$ in
\cubiczetaraw , do a lot of integrations by parts and
drop total derivative terms to obtain
\eqn\cubiczetacompar{\eqalign{
S_3 =
 &\int { 1\over 4 } { \dot \phi^4 \over \dot \rho^4} [ e^{ 3 \rho}
 \dot \zeta^2 \zeta + e^{ \rho}
(\partial \zeta)^2 \zeta ] - { \dot \phi^2
 \over \dot \rho^2} e^{3 \rho} \dot \zeta \partial_i \chi \partial_i
 \zeta +
\cr
 & -{ 1 \over 16} { \dot \phi^6 \over \dot \rho^6} e^{3 \rho}
\dot \zeta^2 \zeta +  {\dot \phi^2 \over \dot \rho^2 } e^{3 \rho}
\dot \zeta \zeta^2 {d \over dt} \left[{ 1 \over 2}  {\ddot \phi \over
\dot \phi \dot \rho } + { 1 \over 4} {\dot \phi^2 \over \dot \rho^2}
\right]
+ { 1 \over 4} { \dot \phi^2 \over
\dot \rho^2 } e^{ 3 \rho} \partial_i \partial_j \chi
\partial_i \partial_j \chi  \zeta
\cr
 ~~    &~~ + f(\zeta) \left.
{ \delta L \over \delta \zeta } \right|_1
}}
where the first line indicates the
leading order terms, which are of order $\epsilon^2$ as expected.
Note that $\chi$  is of order $\epsilon$, see  \first.
The second line indicates higher order terms in the slow roll expansion.
Finally the third line denotes terms that are proportional to the
first order equations of motion \classeq .
They can be removed by
performing a field redefinition of the form
\eqn\zetafieldredef{\eqalign{
\zeta = &  \zeta_n - f(\zeta_n)
\cr
\zeta = & \zeta_n +
 { 1 \over 2}
{ \ddot  \phi \over \dot \phi \dot \rho}
\zeta^2 + { 1 \over 4} { \dot \phi^2 \over \dot \rho^2 }
\zeta^2
+
\cr ~&~
+ { 1 \over \dot \rho} \dot \zeta \zeta
-{1 \over 4} { e^{ -2 \rho} \over \dot \rho^2 } (\partial \zeta)^2
+ { 1\over 4}
{ e^{ -2 \rho} \over \dot \rho^2 } \partial^{-2} \partial_i
\partial_j ( \partial_i \zeta \partial_j \zeta)
+ {1 \over 2} { 1 \over \dot \rho} \partial_i \chi \partial_i \zeta
- { 1 \over 2} { 1\over \dot \rho} \partial^{-2} \partial_i
\partial_j ( \partial_i
 \chi \partial_j \zeta)
% - { 1 \over 4} { \dot \phi^2 \over \dot \rho^3}
%\dot \zeta \zeta +
%\cr
%& ~~ { 1 \over 4} e^{- 2 \rho} \partial^{-2} \partial_i \partial_j
%\delta^2 h^t_{ij}
}}
where  we have written  the explicit expression for $f$\foot{
Note that it does not matter whether  set $\zeta$ or $\zeta_n$ in the
quadratic terms.}.
After this field redefinition the action in terms of $\zeta_n$ is just
the first two lines of \cubiczetacompar .

By comparing the field redefinition \zetafieldredef\ with \valueofT ,
\changeofgauge , \formofzeta , we find that
\eqn\rela{
\zeta_n = - { \dot \rho \over \dot \phi} \varphi
}
with no quadratic correction.
This provides a consistency check on our computations since
it is clear that the actions for $\zeta_n$ and $\varphi$ have
the same form to leading order in slow roll (the first line
of \actthr\ should be compared with the first line of \cubiczetacompar )
and indeed the $\zeta$ and $\zeta_n$ are related as we expect by
the corresponding change of gauge.
The agreement between the two forms of the action should persist to
all others in slow roll but we did not verify it explicitly.

It should be noted that
the field redefinition \zetafieldredef\
does indeed matter
for our computation since we are interested in computing
expectation values of $\zeta$ and not of $\zeta_n$. The reason is that
$\zeta$ is the variable that stays constant outside the horizon while
$\zeta_n$ does not. This last fact  follows  from
the fact that $\zeta$ is constant and the form of \zetafieldredef\
where  some of the
coefficients of the quadratic terms  are time dependent. We can
also see from the action \cubiczetacompar\ that the second line
involves a term with only one time derivative on $\zeta_n$. This
term leads to evolution of $\zeta_n$ outside the horizon
 consistent with what is expected
from \zetafieldredef . Note that only the terms in the  middle
line of \zetafieldredef\ are important outside the horizon.

An easy  way to perform the computation is
to compute using the $\varphi$ or $\zeta_n$ variables up to
a few Hubble times outside horizon exit of the relevant modes and
then change variables to the $\zeta$ variables where it is
clear that there is no evolution outside the horizon.

In order to  perform the computation of the three point function
we will use
a  variable $ \zeta_c$ defined through
\eqn\compvar{
\zeta = \zeta_c  + {1 \over 2} { \ddot \phi \over
\dot \phi \dot \rho} \zeta_c^2  + { 1 \over 8} { \dot \phi^2
\over \dot \rho^2 } \zeta_c^2 + { 1 \over 4} { \dot \phi^2
\over \dot \rho^2 } \partial^{-2} ( \zeta_c \partial^2 \zeta_c)
+  \cdots
}
where the dots indicate terms that vanish outside the horizon  or are
higher order in the slow roll parameters.
In terms of $\zeta_c$ the action becomes
\eqn\actionuse{
S_3  =
%d_t( - { 1\over 2}{ d_t^2 \varphi \dot \varphi \over \dot \rho^3}
%e^{3 \rho } \dot \zeta \zeta^2) -
%d_t ( { 3 \over 8} { \dot \varphi^4 \over \dot \rho^4}
%e^{3 \rho} \dot \zeta \zeta^2 )
%+
\int  { \dot \phi^4 \over \dot \rho^4}  e^{ 5 \rho} \dot \rho
\dot \zeta^2_c \partial^{-2} \dot \zeta_c  + \dots
}
where the dots again indicate terms of higher order in slow roll.
The last term in \compvar\ comes from terms proportional to the equations
of motion that arise when we integrate
 by parts the first line in \cubiczetacompar\ in order
to get \actionuse .

\subsec{ Cubic term for two scalars and a graviton}

We start with the computation in the gauge \firstg
\eqn\actgsca{
S = \int - 2{e^{\rho} \over \dot \rho} \gamma_{ij} \partial_i \dot \zeta
\partial_j \zeta
- e^{\rho} \gamma_{ij} \partial_i \zeta \partial_j \zeta
- { 1\over 2} e^{ 3 \rho} ( 3 \zeta - { \dot \zeta \over \dot
\rho}) \dot \gamma_{ij} \partial_i \partial_j \psi
+ {1 \over 2} e^{3 \rho} \partial_l \gamma_{ij} \partial_i \partial_j \psi
\partial_l \psi
}

Again, it is easiest to understand the dependence of the action
on the slow roll parameter by computing the action in the
gauge \secondg , where the leading contribution comes from
\eqn\actigrv{
S =  \int { 1\over 2} \tilde \gamma_{ij} \partial_i \varphi \partial_j \varphi
+ \cdots
}
where the dots indicate terms that are of higher order in slow roll.
We can then conclude that, despite appearances, \actgsca\ should
be of order $\epsilon$.

Indeed, after some integrations by parts, we find that \actgsca\ becomes
\eqn\actiongrscalar{\eqalign{
S= &\int { 1 \over 2} { \dot \phi^2 \over \dot \rho^2 } e^\rho
 \gamma_{ij} \partial_i \zeta \partial_i \zeta +
\cr
+& { 1 \over 4} e^{ 3 \rho} \partial^2 \gamma_{ij} \partial_i \chi
\partial_j \chi + { 1 \over 4} { \dot \phi^2 \over
\dot \rho^2} e^{ 3 \rho} \dot \gamma_{ij} \partial_i \zeta
\partial_j \chi
 + \hat f(\zeta ,\gamma) \left. { \delta L \over \delta \zeta}\right|_1 +
\hat f_{ij}(\zeta) \left. { \delta L \over \delta \gamma_{ij} }\right|_1
}}
where the first line indicates the leading order term in slow roll,
which indeed has the same slow roll dependence as \actigrv , once
\rela\ is taken into account. The second line contains a
higher order term in the slow roll approximation as well as well
as terms proportional to the equations of motion. These terms
can be removed by field redefinitions which we give explicitly in
appendix A.
These have  the form that we expect
from changing the gauge from \firstg\ to \secondg .
It turns out that these field redefinitions are not important after
horizon crossing and hence are not important for our computation.

\subsec{Cubic term for two gravitons and a scalar}

Let us first do the computation in the gauge \firstg . By direct
substitution in the action and after some integrations by parts
and dropping total derivatives we get
\eqn\twogrscal{\eqalign{
S =& \int { 1 \over 16} {\dot \phi^2 \over \dot \rho^2}
[ e^{3 \rho} \zeta \dot \gamma_{ij} \dot \gamma_{ij}
+ e^\rho \zeta \partial_l \gamma_{ij} \partial_l \gamma_{ij} ]
- { 1 \over 4} e^{3\rho} \dot \gamma_{ij} \partial_l \gamma_{ij}
\partial_l \chi
\cr
 & ~~-  \zeta \dot \gamma_{ij} { \delta L \over \delta \gamma_{ij}} +
 \cdots
}}
As usual the second line can be removed by a field redefinition.
This field redefinition is the same one that we have to use
to go from the gauge \firstg\ to the gauge \secondg, as is discussed
in more detail in appendix A. When we do the computation in
\secondg\ we get directly the first line of \twogrscal , after
taking into account \rela .
It is
curious  that the form of the first line in \twogrscal\ is rather
similar to the first line in \cubiczetacompar . As we did in that
case, in order to perform the computation it is convenient to do
the further field redefinition
\eqn\fieldredgrs{
\zeta = \zeta_c - { 1 \over 32} \gamma_{ij}
\gamma_{ij}  + { 1 \over 16} \partial^{-2} ( \gamma_{ij} \partial^2
\gamma_{ij} )
+ \cdots
}
where the dots indicate terms that vanish outside the horizon.
Then the action becomes
\eqn\actiontwogrscal{
S = \int { 1 \over 4} { \dot \phi^2 \over \dot \rho^2}
\dot \rho e^{ 5 \rho} \dot \gamma_{ij} \dot \gamma_{ij}
\partial^{-2} \dot \zeta_c
+ \cdots}
where the dots indicates terms that are higher order in the slow roll
approximation.

\subsec{Cubic term for three gravitons}

The computation of the term involving three gravitons is
the same in the gauge \firstg\ or the gauge  \secondg , since after setting the scalar
to zero we are changing the metric in precisely the same way.
We note that the only terms in the action that contribute come
from
\eqn\actterm{
S = { 1 \over 2} \int e^{ 2 \rho} ( \hat R +  E^i_j  E^j_i )
}
This has the same form as the result we would have obtained if
we had done the computation in flat space, except for the factor
of $e^{2 \rho}$.
The variable $\hat h_{ij}$ was defined in terms of $\gamma$ in
such a way that there is no  cubic term involving two time derivatives.
This implies that when we integrate by parts we will not need to
use the equations of motion and therefore there will not be
any field redefinitions. In flat space we know from the
form of the scattering amplitudes that we can reduce the
effective vertex to a term involving only spatial derivatives.
We give
its explicit form in the next section.

\newsec{Computation of three point functions}

In this section we compute the three point functions using each of
the interaction lagrangians that we found above.
Before describing the computation let  us make a couple of
general remarks.

First let us notice that we are computing an expectation value and
not a transition amplitude. We want to compute, in the interaction
picture, \eqn\want{ \langle \zeta^3(t) \rangle = \langle
U^{-1}_{int}  \zeta^3(t) U_{int}(t,t_0) \rangle ~,~~~~~~~~ U_{int}
= T \  e^{ - i \int^t_{t_0} H_{int}(t') dt'}   } where $t_0$ is some
early time\foot{So we are not computing $\langle T \zeta^3 e^{-i
\int_{-\infty}^{\infty} H_{int} } \rangle$ which is what we
compute when we have scattering amplitudes in mind. When we
compute scattering amplitudes field redefinitions  do not change
the answer. In our  computation they do. }, and $T$ denotes a time ordering.
 We have suppressed the spatial
dependence.
To first order this gives
\eqn\wantfirst{
\langle \zeta^3(t) \rangle = -i \int_{t_0}^t dt'\langle [\zeta^3(t) ,
H_{int}(t') ] \rangle
}
For the cubic terms $H_{int} = - L_{int}$ after we remove all terms
proportional to the equations of motion by a field redefinition.

Our second point is to
note that we want to compute the expectation value in
the vacuum of the interacting theory, not the vacuum of the free
theory. When we do  computations in Minkowski space we also
have to take this into account. This can be automatically
taken into account by deforming the $t'$ integration contour
so that it includes some evolution in euclidean time which
projects on to the true vacuum. Fortunately in this case
we can apply a similar procedure to select the vacuum. The
basic reason is that at early times the physical wavelength
is very small and we feel in  Minkowski space, so we want
the vacuum for these high energy modes to be what it is in
the interacting theory in this approximately Minkowski space.
In de-Sitter space this is the
 Hartle Hawking prescription
for the vacuum \hh .
In practice this will translate into
a choice of contour for the integral in \wantfirst .\foot{
Other choices of vacua in de-Sitter space were discussed in
\othervacua\ . In inflation the admixture of these other
vacua is expected to be
small \othersmall\  and the leading contribution to the three point
function comes from the usual vacuum.}
The evaluation of the integral in \wantfirst\ reduces to
an integration of the cubic action evaluated on the classical
solutions of \classeq \foot{
This is true only after performing  field redefinitions to eliminate
terms proportional to the equations of motion. Otherwise we need to take
into account total derivative terms that arise
when we go from \cubiczetaraw\
to \cubiczetacompar ,  for example. }. Since we do not have the solutions for
a general potential it is useful to choose a method that
minimizes the errors in the approximate evaluation of the integrals.
These errors are minimized if we split the integrals in
\wantfirst\ as an integral over  the region outside the horizon, the
region around horizon crossing and the region deep inside the horizon.
In the last region the fields oscillate rapidly and after our
continuation to Euclidean space there is no contribution. In the
region near horizon crossing we approximate the solutions
by those of de-Sitter space \classsolds\ and we use the action
in the form that shows the leading slow roll dependence, such
as in \actionuse ,  \actiongrscalar\  or \actiontwogrscal .
After we exit the horizon
we know that $\zeta$ and $\gamma$ are constant so we switch to
those variables. This is taken into account by the field redefinitions
we talked about. Then in the region well outside the horizon
the fields are constant and the integral \wantfirst\ vanishes when
we do the computation in
 the $\zeta , ~ \gamma$
variables.
Below we proceed with this computation for the various cases.
Due to momentum conservation there
are basically two distinct kinematic configurations, the
$k_i$ can be all of the same order of magnitude of one of
the $k_i$ is much smaller than the other two.
We will consider these two cases separately.

\subsec{Three scalars correlator}

Note that if we have a field redefinition of the schematic form
$\zeta = \zeta_c + \lambda \zeta_c^2$ then the correlation function
will contain two terms
\eqn\formcontr{
\langle \zeta(x_1) \zeta(x_2) \zeta(x_3) \rangle =
\langle \zeta_c(x_1) \zeta_c(x_2) \zeta_c(x_3) \rangle +
2 \lambda \left[ \langle  \zeta(x_1) \zeta(x_2)\rangle
 \langle  \zeta(x_1) \zeta(x_3)\rangle  + {\rm cyclic} \right]
}
The first term is computed by \actionuse . The second comes from
the field redefinition \compvar .
 By performing
different field redefinitions we can reshuffle the contributions
between these two terms.
Let us first compute the contribution from the
action \actionuse . As we explained above we evaluate this term
in de-Sitter space with parameters corresponding to those of
horizon exit.
In de-Sitter space
 the contribution of an action of
the form  \actionuse\ has the form
\eqn\threedes{\eqalign{
\langle \zeta_c \zeta_c \zeta_c   \rangle = & ( 2 \pi)^3 \delta^3(\sum \vec k_i)
{ 1 \over \prod (2 k_i^3) } { \dot \rho_*^6 \over \dot \phi^2_*}
 i \int_{-\infty}^0
{d \eta } k_1^2 k_2^2 e^{i k_t \eta}
+ {\rm permutations} + c.c.  =
\cr
=& ( 2 \pi)^3 \delta^3(\sum \vec k_i)
{ 1 \over \prod (2 k_i^3) } { \dot \rho_*^6 \over \dot \phi^2_*}
 4 { \sum_{i>j} k_i^2 k_j^2 \over
k_t } ~,~~~~~~~~~~~~~~{ \rm with} ~~ k_t = k_1 + k_2 + k_3
}}
Note that $k_i \equiv  |\vec k_i|$.
The choice of integration  contour in \threedes\
is such that the oscillating piece
in the exponent becomes exponentially decreasing. In other words we
change  $ \eta \to \eta + i \epsilon|\eta| $ for large $|\eta|$.
 This choice
of contour is the one corresponding to the standard vacuum of the
interacting theory.

After adding the contribution of the field redefinitions we
get the final result for the three point function
\eqn\finalres{
\langle \zeta_{\vec k_1} \zeta_{ \vec k_2} \zeta_{\vec k_3} \rangle =
( 2 \pi)^3 \delta^3 ( \sum \vec k_i) { \dot \rho^4_* \over
\dot \phi^4_* }{ H^4_* \over M_{pl}^4 } { 1
\over \prod_i ( 2 k_i^3) } { \cal A}_*
}
where the star indicates evaluation at horizon crossing and
\eqn\resultfora{
{ \cal A} =2 { \ddot \phi_* \over
\dot \phi_* \dot \rho_*} \sum_i k_i^3  + { \dot \phi^2_* \over
\dot \rho^2_*}
\left[{ 1 \over 2}  \sum_i k_i^3  + { 1 \over 2} \sum_{i \not = j}
k_i k_j^2 +
4 { \sum_{i>j} k_i^2 k_j^2 \over
k_t } \right]
}
In writing \finalres\ and \resultfora\
we have assumed that all $k$'s are of the same order of magnitude
so that the moment of horizon crossing does not differ too much
between the different modes. Due to momentum conservation
the other possibility is that one
of the $k$'s is much smaller than the other two and these last
two would be of the same order of magnitude. So we consider
the configuration $k_3 \ll k_1 \sim k_2 $.
The mode labeled by $k_3$ crosses the horizon much earlier than the
other modes. By  the time that
$k_{1,2}$ cross the horizon  $\zeta_3$ is
constant. The only effect of the $\zeta_3$ fluctuation will
be to make the comoving scales $k_{1,2}$ cross the horizon
at a slightly earlier  time $\delta t_* = - \zeta_3/\dot \rho_{*}$.
This will produce a change in the fluctuations with momenta $k_{1,2}$
due to the time dependence of the slow roll factors in \zetaapprox .
In conclusion we obtain
\eqn\simpleget{\eqalign{
\langle \zeta_{\vec k_1} \zeta_{\vec k_2} \zeta_{\vec k_3} \rangle \sim
 & -  \langle \zeta_{\vec k_3} \zeta_{- \vec k_3} \rangle'
 { 1 \over \dot \rho_*}
 {d \over dt_*}  \langle \zeta_{\vec k_1} \zeta_{\vec k_2}\rangle
\cr
\sim & - n_{s*} \langle \zeta_{k_3 } \zeta_{- \vec k_3} \rangle
\langle \zeta_{\vec k_1 } \zeta_{\vec k_2} \rangle
\cr
\sim &
( 2 \pi)^3 \delta^3 ( \sum_i \vec k_i)
 { \dot \rho^4_* \over
\dot \phi^2_* M_{pl}^2}{ \dot \rho^4_{*'} \over
\dot \phi^2_{*'} M_{pl}^2}
{ 1 \over
2 k_1^3 2 k_3^3 } 2 ( { \ddot \phi_* \over \dot \rho_* \dot \phi_*} +
{\dot \phi^2_* \over \dot \rho^2_* })
}}
where now * indicates the moment that $k_{1},~k_2$ cross the
horizon and $*'$ indicates the time when $k_3$ crosses the
horizon (which is earlier). In the second line we point out explicitly that
this two point function involves the tilt of the scalar spectrum
$n_{s*}$ \defns , evaluated at $t_*$.
The prime in the first two lines of \simpleget\
means that we  omit the   factor  $(2\pi)^3 \delta( \vec 0 )$.
It can be checked that \finalres , with \resultfora\
goes over to \simpleget\ in the overlapping region of validity which
is when $k_3$ is small but not so small to change the slow roll parameters
appreciably. The first two lines in \simpleget\ are valid to all orders
in slow roll parameters in the regime $k_3 \ll k_{1,2}$.

Our result \finalres\ is of the same order of magnitude as in
\gangui \acquaviva , but the $k$ dependence as well as the precise
numerical coefficients are different. The reason is that
\acquaviva\ considers only effects due to non-linear evolution but
do not consider the change in vacuum. Both of these
effects are of the same order of magnitude so they should be both
included and are intimately linked. Our result obeys
consistency condition explained in the above paragraph while that in
\acquaviva\ does not, presumably because not all relevant effects
were included.

Spergel and Komatsu \ks \ksother \komatsu\
did an extensive analysis of  the measurability
of the three point function. They assumed
that the three point function had the form that
would follow from a field redefinition of the form\foot{
Spergel and Komatsu defined $\Phi = \Phi_g + f_{NL} \Phi_g^2 $.
The factor of $5/3$ arises in the relation between the
gauge invariant newtonian potential $\Phi$ and $\zeta$ during matter
domination, $\zeta = - { 5\over 3} \Phi$. }
\eqn\definitionoffnl{
\zeta = \zeta_g - { 3 \over 5}  f_{NL} \zeta_g^2
}
where $\zeta_g$ is gaussian.
Their analysis can be roughly summarized by
saying  that this would be measurable if $f_{NL }$ is bigger than
around 5. This constraint comes mainly  from cosmic variance if we
assume that we measure the CMB up to the angular scales that
the Planck satellite will measure them. See \komatsu\ for
a detailed discussion of this point.
Our final result \resultfora\ does not have the momentum dependence
 that would
follow from \definitionoffnl . In order for that to be the case
we would need that all terms in \resultfora\ were proportional to
$\sum_i k_i^3$ which is clearly not the case. So we cannot
recast our computation as a computation of $f_{NL}$.

Nevertheless we can define  a $k$-dependent $f_{NL} $ as
\eqn\estimate{
- f_{NL} \sim  { 5 \over 3} { { \cal A} \over (4 \sum_i k_i^3)}
= { 5 \over 12}  \left[ 2 { \ddot \phi_* \over
\dot \phi_* \dot \rho_*} + { \dot \phi^2_* \over
\dot \rho^2_*} ( 2 + f(k) ) \right]=
- { 5 \over 12 } (n_s + f(k) n_t) }
where $f(k)$ has a  range  of values  $ 0  \leq f \leq
{ 5\over 6} $. $f(k)$  is a function of the shape of the triangle
made by $\vec k_i$ and it  goes to zero  when  two sides become
much larger than the third and it becomes 5/6
 when the $\vec k_i$ form an
equilateral triangle.

Assuming that this $k$ dependence of the three point function does
not significantly change the analysis in \ks \ksother \komatsu ,
 we
unfortunately  conclude that
it will not be possible to see this effect purely
from the CMB.
Actually, the discussion of \komatsu\ makes sense for
for $f_{NL} > 1$ where we can neglect the non-linearities in the
gravitational evolution after horizon reentry, some of these
effects were discussed by \paynecaroll \luo \starobinskytwo .
 In other words, to
measure $f_{NL} <1$,  one
has to include the leading non-linear effects
in the whole evolution until we measure the temperature of
the CMB.

\subsec{Two scalars and a graviton  correlator}

This computation is rather similar to the one we
did above so we will not repeat all the details. Let us
note that there is no field redefinition that is
important at late times so that we only need to evaluate the
integral that arises from  the interaction term in
the first line of \actiongrscalar .

This gives
\eqn\grscalcorr{
\langle \gamma^{s}_{\vec k_1} \zeta_{\vec k_2} \zeta_{\vec k_3 } \rangle
= (2 \pi)^3 \delta^3( \sum \vec k_i )  { 1 \over
\prod (2 k_i^3) } { \dot \rho_*^4 \over M_{pl}^4} { \dot \rho^2_*
\over \dot \phi^2_* }
{ \epsilon^s_{ij} k^i_2 k_3^j} 4 I
%  ( -k_t + { \sum_{i>j} k_i k_j \over k_t}
%+ {k_1k_2 k_3 \over k_t^2})
}
where the transverse and traceless
  polarization tensor is normalized
to $\epsilon^s_{ij}\epsilon^{s'}_{ij} = 2 \delta_{ss'}$ and
$I$ is
\eqn\integr{\eqalign{
I =& Re\left[-   \int^0_{-\infty} i { d\eta \over \eta^2}
 ( 1 -i  k_1 \eta)( 1 - i  k_2 \eta)( 1 - i  k_3 \eta )
e^{ i k_t \eta } \right]
\cr
I = & -k_t + { \sum_{i>j} k_i k_j \over k_t}
+ {k_1k_2 k_3 \over k_t^2}
}}
The integral in \integr\ diverges at $\eta \to 0$ but the divergence is
purely imaginary so that $I$ is finite with our choice of
 contour\foot{In order to evaluate this integral
it is convenient to note that $Re [- i \int_{-\infty}^0 d\eta \eta^{-2}
(1 - i \eta) e^{i\eta} ] = -1 $ with our contour prescription. }.

The dependence on the  slow roll parameters is  such that the three
$\zeta$ correlator and the $\gamma \zeta^2$ correlator are of the
same order of magnitude. After horizon reentry the amplitude of
the gravitational
waves  decays so that for high $l$ we still expect the three
$\zeta$ correlator to dominate.

Let us now consider the correlation function in the limit
 $k_1 \ll k_2,~k_3$. When $k_2,k_3$ are about
to cross the horizon the gravity wave with momentum $k_1$ is already
frozen so that the fluctuations of $\zeta$ will be those that we
expect in this deformed geometry. The main effect of
the deformation is to change $k^2 \to k^2 - \gamma_{ij} k^i k^j $
in the correlation function of two $\zeta$s.
This reasoning leads to
\eqn\corrdiffks{\eqalign{
\langle \gamma^{s}_{\vec k_1} \zeta_{\vec k_2} \zeta_{\vec k_3}
\rangle' \sim &
 - \langle \gamma^s_{\vec k_1}
 \gamma^s_{-\vec k_1}  \rangle
 \epsilon^{s}_{ij} k_2^i k^j_2   { \partial \over \partial k_2^2 }
 \langle \zeta_{\vec k_2} \zeta_{\vec k_3} \rangle
\cr
\sim & (2 \pi)^3 \delta(\sum \vec k_i) { 1 \over 2 k_2^5}{ 1 \over 2 k_1^3}
 { \dot \rho^4_* \over \dot \phi^2_* M_{pl}^2 } {2 \dot \rho_{*'}^2
\over M_{pl}^2 }{ 3\over 2} { \epsilon^s_{ij} k_2^i k_2^j}
}}
where * denotes the time when $k_{2,3}$ cross the horizon while
*' denotes the time when $k_1$ crosses the horizon.

We see that \corrdiffks\ and \grscalcorr\ are consistent in the
overlapping region of validity. This is a consistency check of
the computation.

\subsec{Two gravitons and a scalar correlator}

The evaluation of this correlator using \fieldredgrs\ and
\actiontwogrscal\ is very similar
to the one of the three scalar correlator. We obtain
\eqn\finaltwograv{
\langle \zeta_{\vec k_1} \gamma^{s_2}_{\vec k_2}
 \gamma^{s_3}_{\vec k_3}\rangle =
{ (2 \pi)^3 \delta( \sum \vec k_i)
\over \prod (2 k_i)^3 } { \dot \rho^4 \over M_{pl}^4}
[-{ 1\over 4} k_1^3 +
{ 1 \over 2} k_1 (k_2^2 + k_3^2) + 4 { k_2^2 k_3^2 \over k_t} ]
\epsilon^{s_2}_{ij} \epsilon^{s_3}_{ij}
}

In the case that $k_1 \ll k_{2,3}$ we also find that the
correlation function is given by the derivative of the slow roll
factor in the correlation function of two tensor fluctuations.
We get
\eqn\smallktwogr{\eqalign{
\langle \zeta_{\vec k_1} \gamma^{s_2}_{\vec k_2}& \gamma^{s_3}_{\vec k_3}
\rangle \sim
-  \langle
\zeta_{\vec k_1} \zeta_{- \vec k_1} \rangle'  { 1\over \dot \rho_*}
{ d \over dt_*}
\langle  \gamma^{s_2}_{\vec k_2} \gamma^{s_3}_{\vec k_3} \rangle
\cr
& \sim - n_t
\langle \zeta_{\vec k_1} \zeta_{- \vec k_1} \rangle'
\langle \gamma^{s_2}_{\vec k_2} \gamma^{s_3}_{\vec k_3} \rangle
\cr
&
\sim { (2 \pi)^3 \delta(\sum \vec k_i) }
{ \dot \phi^2_* \over \dot \rho^2_*}{ \dot \rho^2_* \over  M_{pl}^2}
{ 2 \delta_{s_2 s_3} \over 2 k_2^2} {\dot \rho^4_{*'}
\over \dot \phi_{*'}^2 M_{pl}^2} { 1 \over 2 k_1^3}
}}
where we have used that $\vec k_2 \sim \vec k_3$ so that
$\epsilon^{s_2}_{ij} \epsilon^{s_3}_{ij} \sim 2 \delta_{s_2s_3} $.
The $*$ indicates horizon crossing for $k_2,~k_3$ and the
$*'$ indicates horizon crossing for $k_1$ which happens earlier.
In the second line we emphasized the dependence of this three point
function on the tilt of the gravity wave spectrum.
We see that \smallktwogr\ agrees with \finaltwograv\ in the overlapping
region of validity.

\subsec{Three graviton  correlator}

The three graviton correlator is  a very similar computation.
The algebra involving polarization tensors is the same as in
flat space so that we can use the flat space result.
We will need to do the same integral as in \integr .
The final result is
\eqn\finalthreegr{
\langle \gamma^{s_1}_{\vec k_1} \gamma^{s_2}_{\vec k_2}
\gamma^{s_3}_{ \vec k_3 }
\rangle
= (2 \pi)^3 \delta^3 ( \sum \vec k_i)  { \dot \rho^4_* \over M_{pl}^4}
{ 1 \over \prod_i (2 k_i^3)}
(-4) (\epsilon^{s_1}_{i i'}\epsilon^{s_2}_{j j'} \epsilon^{s_3}_{l l'}
 t_{ijl} t_{i'j'l'} ) I
}
where $I$ is given in \integr , and $t_{ijk}$ is given by the
flat space formula (see for example \polchinski )
\eqn\formoft{
t_{ijl} =  k_2^i \delta_{jl} + k_3^j \delta_{il} + k_1^l \delta_{ij}
}

We can compute this in the limit $k_1 \ll k_{2,3}$ in a way similar to
what we did for the case of a graviton and two scalars
\eqn\finalthrsmall{
\langle \gamma^{s_1}_{\vec k_1} \gamma^{s_2}_{\vec k_2}
\gamma^{s_3}_{ \vec k_3 }
\rangle
= (2 \pi)^3 \delta^3 ( \sum \vec k_i)
{ 2  \delta_{s_2 s_3} \dot \rho^2_* \over M_{pl}^2}
{ 1 \over 2 k_2^5}{ 2 \dot  \rho^2_{*'} \over M_{pl}^2}{ 1 \over 2 k_1^3}
{ 3 \over 2} \epsilon^{s_1}_{ij} k_2^i k_2^j
}
which indeed agrees with \finalthreegr\ in the overlapping region of
validity.

\newsec{ Remarks on AdS/CFT and dS/CFT }

\subsec{ AdS/CFT}

The computation that we did above was done with inflation in
mind, but the same mathematical structure arises if one
considers a single scalar field with a negative potential.
In the slow roll case, the background will be a slightly deformed
anti-de-Sitter space. This can be understood as a
slightly deformed conformal field theory. In other words, a
non-conformal field theory which is almost conformal.
An incomplete list of references where situations of this
sort were considered is
\porrati \gubserctheorem
\frolov \freedman \skenderisone \skenderistwo . Here we
just mention a few results that are relevant for us,
for a review see \skenderisreview .
The variables $\gamma^s $ that we used above  are
associated to the traceless components of the stress tensor
while the
variable $\zeta$ is associated to the trace of the stress tensor.
More precisely, we have a coupling of the form
$\int { dk^3 \over (2 \pi)^3 }[ 2 \zeta_{- \vec k} T^i_i(\vec k) +
2 \gamma^s_{-\vec k}  T^s( \vec k ) ]$, where $T^s$ is defined
by an expression similar to \expans , with $\gamma \to T$.
The fact that the definition of the scalar mode depends
on the gauge is translated into the fact that in a field
theory with a scale we can either change the dimensionfull coupling
constant or we can change the overall scale in the metric.
It is common to fix the coupling and change the metric, which
then relates $\zeta $ to the trace of the stress tensor. Alternatively
we can fix the metric and change the coupling constant. In the field
theory we do not have two independent operators, we have only one
operator related by the equation
\eqn\tracte{
2 T_i^i = \beta_\lambda  {\cal O}
}
where $\beta_\lambda$ is the beta function for the coupling $\lambda$
which appears in the field theory Lagrangian in front of the
non-marginal operator  as  $ \int \lambda {\cal O}$.
The operator ${\cal O}$ is the one coupling to $\phi$ and the
operator $2 T^i_i$ couples to $\zeta$. The factor of slow roll
that relates the correlators of $\zeta$ and $\phi$ is
precisely the factor $\beta_\lambda$
appearing above \larsen .

From the computations in the previous sections
 we can also compute
the correlation function of stress tensors and trace of the
stress tensor in non-conformal theories. Depending on whether
the slow roll approximation is valid or not we would need to use
different formulae in those sections.

Two point functions of the trace of the stress tensor
 were considered in the $AdS$ context in  \freedman \frolov
 \skenderisone \skenderistwo . The derivation of the effective
 action for the corresponding field in $AdS$ identical to the
 one in the $dS$ context.
Similarly, computations of three point functions in AdS
can be done by  performing minor modifications to the above formulae.
We will be more explicit below.

Now we will review the $AdS_4$ computation (see \review\ for a review)
so that we can
contrast it  clearly to the $dS_4$ computation.

Let us consider a canonically normalized scalar field in
Euclidean anti-Sitter space ($EAdS_4$) which is the same as hyperbolic
space.
The action is
\eqn\actn{
S =  R^2_{AdS} \int { dz \over z^2 } { 1\over 2}
 [ ( \partial_z f)^2 + (\partial f)^2 ]
}
In order to do computations it will be necessary  to consider
classical solutions which go to zero for large $z$ and obey
prescribed boundary conditions at $z = z_c$. In momentum
space these are
\eqn\clsolmom{
f_{\vec k} = f_{\vec k}^0
{ ( 1 + k z)e^{ - k z} \over ( 1 + k z_c) e^{ - k z_c} }
~,~~~~~~~~~k = |\vec k| }
where $f^0_{\vec k}$ is the boundary condition we impose at $z=z_c$.
One should then compute  the action for this solution as a function
of the boundary conditions. Inserting \clsolmom\ into \actn , integrating
by parts and using the equations of motion we get
\eqn\actionads{\eqalign{
-S = & \int {d^3k \over (2 \pi)^3 }
{ 1 \over 2}  R^2_{AdS}  f_{-\vec k}^0 { 1 \over z^2_c }
\left. { d f_{\vec k} \over dz}\right|_{ z=  z_c } =
-\int {d^3k \over (2 \pi)^3 }{ 1 \over 2} R^2_{AdS}  f_{- \vec k}^0 f_{ \vec k}^0
{ k^2 \over  z_c ( 1 + k  z_c) }
\cr
\sim &- \int {d^3k \over (2 \pi)^3 } { 1 \over 2}  R^2_{AdS}  f_{- \vec k}^0 f_{ \vec k}^0
 [ { k^2  \over  z_c } - k^3 + \cdots ]
}}
where the dots indicate terms of higher order in $z_c$.
The term divergent in $ z_c$  is local in position space\foot{
It is proportional to $ { 1 \over z_c}
\int dx^3  { 1 \over 2} (\partial f^0)^2 $.}
and it is viewed as a  divergence in the CFT which should be  subtracted by
a local counterterm.
The term independent of
$z_c$ is non-local and gives rise to the two point function
\eqn\twopoint{
\langle { \cal O}(\vec k) { \cal O}(\vec k') \rangle_{EAdS}
= \left. { \delta^2 Z \over \delta f^0_{\vec k} \delta f^0_{\vec k'} }
\right|_{f^0=0}  \sim
 ( 2 \pi)^3 \delta( \vec k + \vec k'){ R^2_{AdS} }   k^3
}
Where $Z$ is the partition function of the Euclidean CFT which
is approximated by $Z \sim  e^{ - S_{cl}}$, with $S$ in \actionads .

\subsec{dS-CFT}

The dS/CFT was proposed
\stromingerds \wittends\
in analogy with AdS/CFT \jm \wittenhol \gkp .
The dS/CFT postulates that the wavefunction of a universe which
is asymptotically de-Sitter space can be computed in terms
of a conformal field theory.
More precisely, we have the formula
\eqn\dscft{
\Psi[g] =  Z[g]
}
where the left hand side is the wavefunction of the universe
for given three metric and the right hand side is the
partition function of some dual conformal field theory.
Actually the left hand side has rapidly oscillating pieces which
can be expressed as local functions of the metric. We discard these
pieces since they have the interpretation of local counterterms in
the CFT.
Here we are thinking of de-Sitter in flat slices (or Poincare coordinates)
and we are
imagining that all fields start in their life in the Bunch-Davies
vacuum. This determines the wavefunction $\Psi$, at least in
the context of perturbation theory. If we were considering global
de-Sitter space then our discussion would be valid in a small patch
in the future where it can be approximated by the Poincare patch
and the memory of the particular state that could have come from
the far past is lost\foot{ The information of the state coming from the
asymptotic past in global dS  is contained on modes whose
angular momenta, $l$,  on the sphere is fixed, assuming the evolution is
non-singular and in the context of perturbation theory.
 On the other hand, we are focusing on
modes with $l \gg 1$ when we look at the Poincare patch. }.
 This point of view follows
simply from the discussion in \wittends\ in analogy with the
standard discussion in Euclidean AdS where the same formula \dscft\
is valid\foot{In  AdS/CFT formula \dscft\ arises in the Euclidean
context when we think of Euclidean time as the direction perpendicular
to the boundary. $\Psi$ can then be interpreted also as the
 Hartle-Hawking
wave function \hh .
See \wittentop \periwal \dbvv\
 for more on this point of view.}.
Nobody has found a concrete example of this duality and there
are some  suspicions that such a duality should not exist
\susskind .
All we will do here is to do some computations on the gravity side
in order to get some insight on the properties that this hypothetical
CFT should have.
If an example were found, then it would be a more powerful
way of computing the wavefunction that semiclassical physics
in de-Sitter or nearly de-Sitter space.
Note that an observer living in eternal de-Sitter space will not
be able to measure two point correlators such as \resgr\
 or the wavefunction
\dscft\ which involves distances much larger than the Hubble scale.
Only so called ``metaobservers'' can measure these \wittends .
 On the other
hand if the universe is approximately de-Sitter for a while and then
inflation ends and we go over to a radiation or matter dominated
universe  then these correlation functions become observable.
In fact, we are metaobservers of the early inflationary
epoch \danielson .

In \stromingerds \otherds\
 the relation between CFT operators and fields in the
bulk was explored and various ways of defining operators were
considered. It was found that given a scalar field in the bulk one
could define two operators with two conformal dimensions differing by
$\Delta_+ - \Delta_- = d$ where $d $ is the dimension of the CFT.
If the field we are considering in the bulk is the metric then
it is clear that the corresponding operator is the stress tensor
and it should have dimension $d$.
Indeed we will see that this agrees precisely
with what we expect from the prescription \dscft .
Below we explain more precisely how this computation is related
both to the inflationary computation \resgr\  and the corresponding
EAdS computation.

The first step is to compute the wavefunction as a function of
a small fluctuation in a massless scalar field
$f$. Since $f$ is a free field, which is
a collection of harmonic oscillators, all we need to do is to
compute the wavefunction for these harmonic oscillators.
We want to compute the Schroedinger picture
 wavefunction at some time $\eta_c$ as
a function of the amplitude of the field $f$.
The wavefunction is  given by a sum over all  paths
ending with amplitude $f$ and starting at the appropriate vacuum
state. Since the action is quadratic this sum reduces to evaluating
the action on the appropriate classical solution.
We choose the standard Euclidean (Bunch-Davies)
vacuum for the fields at
early times.
The classical solution obeying the appropriate boundary conditions
is
\eqn\classsol{
f = f^0_{  \vec k} {  ( 1 - i k \eta) e^{ i k \eta }  \over
( 1 - i k \eta_c) e^{ i k \eta_c} }
}
The boundary conditions at large $\eta$ are the ones that
correspond to the statement that the oscillator is in its ground
state, which can be defined adiabatically at early times.
The condition is that the field should behaves as $e^{ i k \eta}$
for $|\eta| \to \infty $.
Note that $f_{ -\vec k} \not = f_{ \vec k}^*$ since the boundary
condition we are imposing at early times is not a real condition
on the field $f(\eta, x)$ \foot{  There is nothing wrong in considering
a complex solution since all we
are doing is to evaluate a functional integral by a saddle point
approximation.}.
This is one of the many ways
 to think about the harmonic oscillator
wavefunction.
When we evaluate the classical action on this solution we get
\eqn\actlor{\eqalign{
iS =  &i \int {d^3k \over (2 \pi)^3 } {1 \over 2}  R^2_{dS}  { 1 \over \eta^2_c}
f^0_{ -\vec k} \partial_\eta f_{ \vec k} |_{\eta = \eta_c}
= i\int {d^3k \over (2 \pi)^3 } {1 \over 2} R^2_{dS} { k^2 \over \eta_c ( 1 - i k \eta_c )}
f^0_{ -\vec k}f^0_{ \vec k} \cr
\sim & \int {d^3k \over (2 \pi)^3 } {1 \over 2}  R^2_{dS}
[  i {k^2 \over \eta_c }
 - k^3 + \cdots ] f^0_{ - \vec k}  f^0_{\vec k}
}}
Note that we are dropping an oscillatory piece at $|\eta| \to \infty$
which is equivalent to slightly changing the contour of integration
by $\eta \to \eta + i \epsilon $. This is the standard prescription
for the vacuum state of a harmonic oscillator.

Notice that under
\eqn\analytic{
\eta = i z ~,~~~~~~~ R_{dS} = i R_{AdS}
}
 the formulas
\classsol\ and \actlor\ go into \clsolmom\ and \actionads .
The fact that \classsol\  goes into  \clsolmom\ is intimately related
to the statement
that when the mode has short wavelength it is in the adiabatic vacuum.
A consequence of this fact is that the two point
function computed using $dS_4$ differs by a sign from the
corresponding one in Euclidean $AdS_4$\foot{ In other dimensions
there are  extra $i$s that appears in the relation.}.
More explicitly we have
\eqn\corrfnds{
\langle { \cal O}(\vec k) { \cal O}(\vec k') \rangle_{dS_4}
\equiv
 \left. { \delta^2 Z \over \delta f^0_{\vec k} \delta f^0_{\vec k'} }
\right|_{f^0=0}  \sim
 ( 2 \pi)^3 \delta( \vec k + \vec k'){ R^2_{dS} } (-  k^3)
}
We can easily check that this is the analytically continued version of
\twopoint\ under \analytic .

Now let us understand the relation between the wavefunction computed in
\actlor , which is $\Psi \sim  e^{iS_{cl}} $ and the expectation
values that appeared in our earlier discussion \twods .
Of course, the relation is  that $ \langle f^2 \rangle =
\int { \cal D} f f^2  |\Psi(f)|^2 $.
We see that only the real piece in $iS$ contributes. This has
a finite limit  at late times. The divergent pieces in
\actlor\  are
all imaginary and do not contribute to the expectation value.
 The functional integration over
$f$ gives again \twods . There is a crucial factor of 2 that
comes from the square of the wavefunction, so  that the
relation between \twods\ and \corrfnds\
is not a Legendre transform.
% This prescription for computing the
% two point functions seems to differ from that given in \stromingerds
% \otherds , but is closely related to the discussion in \wittends .

Our previous discussion focused on a scalar field and its
corresponding operator ${\cal O}$. All that we have
said above translates very simply for the traceless part
of the metric and the traceless part of the stress tensor, since
at the linearized level the action for the graviton in
the traceless transverse gauge  reduces
to the action of a scalar field \gravwv \expans .
We are defining the stress tensor operator as
\eqn\defsten{
T_{i j}(x) \equiv { \delta  Z[h] \over \sqrt{h} \delta  h^{ij}(x) }
= { \delta  \Psi[h] \over \sqrt{h} \delta  h^{ij}(x) }
}
which is a standard definition for a Euclidean field theory.\foot{
One might want to define it with an $i$ so that
$ T_{ jl} \equiv i { \delta Z[h] \over \sqrt{h} \delta  h^{jl} }$. This
definition might be natural given that  the counterterms
(which represent the leading dependence of the wavefunction)
are purely imaginary. In any case,  it is trivial to go between
both definitions. }
In this case the divergent term in \actlor\ can be rewritten as
$  - i{ 1 \over 2 \eta_c} \int d^3 x \sqrt{h} R^{(3)} $.
Note that there is a factor of $i$.
We want
to remove this by a counterterm in the action of the Euclidean
CFT. These factors of $i$ are related to the fact that the renormalization
group transformation
in the CFT should be appropriately unitary since this  RG transformation
 corresponds,  in the context of perturbation theory, to unitary
 evolution of the wavefunction in the bulk.
If we define the central charge of the CFT in terms
of the two point function of the stress tensor we get a negative
answer. This negative answer has a simple qualitative explanation.
We know that the wavefunction in terms of small fluctuations is
bounded, in the sense that it is of the form $e^{ - \alpha |f|^2} $
with $\alpha$ positive, since each mode is a harmonic oscillator
with positive frequency. This sign implies a negative
sign for the two point function of the stress tensor.
Similarly the trace of the stress tensor is related to the derivative
of the wavefunction
with respect to  $\zeta $.

After we understood the relation between two point functions of
operators and expectation values of the corresponding fluctuations
we can similarly understand the relation between three point functions.
The wavefunction has the form
\eqn\wavf{\eqalign{
\Psi = & Exp\left[ { 1 \over 2} \int d^3x d^3x' \langle {\cal O}(x)
{\cal O}(x') \rangle f(x) f(x') + \right.
\cr
& ~~ \left.
 { 1 \over 6} \int d^3x d^3x' d^3 x''\langle {\cal O}(x)
{\cal O}(x'){\cal O}(x'') \rangle f(x) f(x') f(x'') \right]
}}
where we  emphasized that derivatives of $\Psi$
 give correlation functions for the corresponding
operators. The expectation values in momentum space
 are  related by
\eqn\expectval{\eqalign{
\langle f_{\vec k} f_{- \vec k} \rangle' =
 &- { 1 \over 2 Re \langle {\cal O}_{\vec k}{\cal O}_{-\vec k} \rangle' }
\cr
\langle f_{\vec k_1}f_{\vec k_2}f_{\vec k_3} \rangle'=
 & { 2 Re \langle  {\cal O}_{\vec k_1} {\cal O}_{\vec k_2}
{\cal O}_{\vec k_3}
 \rangle'
 \over \prod_i (- 2 Re \langle {\cal O}_{\vec k_i}{\cal O}_{- \vec k_i}
\rangle' )
  }
}}
where the prime means that we dropped a factor of $(2 \pi)^3
\delta( \sum \vec k) $. And $Re$ indicates the real part.
The factors of two  come from the fact
that we are squaring the wavefunction \wavf . Notice that this
explains why $\langle TT \rangle \sim c $ while $\langle \gamma \gamma
\rangle \sim 1/c$ where $ c \sim  - R^2_{dS} M_{pl}^2 $.

Now consider three point functions.
For example,
consider the three point function of the traceless part of the stress
tensor. This
can be computed directly in $dS_4$ by inserting the
classical solutions \classsol\ into the cubic terms in the action.
This gives
\eqn\threeptfn{
\langle T^{s_1}_{\vec k_1} T^{s_2}_{\vec k_2}
T^{s_3}_{ \vec k_3 }
\rangle
= (2 \pi)^3 \delta^3 ( \sum \vec k_i)  {M_{pl}^2 \over \dot \rho^2_* }
( - { 1 \over 32} )
 (\epsilon^{s_1}_{i i'}\epsilon^{s_2}_{j j'} \epsilon^{s_3}_{l l'}
 t_{ijl} t_{i'j'l'} ) I
}
where $I$ is defined in \integr .
The result in $EAdS_4$ is the same as above except for a minus sign, which
can be understood as coming from \analytic . When we perform this
computation
we need to drop a local divergent
term which is proportional to  ${ - i \over 2\eta_c}
 \sqrt{h} R^{(3)}$.
We did not have any divergence in \finalthrsmall\
due to the fact that we were computing the square of the wavefunction
while in \threeptfn\ we are computing the third derivative
of the wavefunction.
Of course, we can compute directly \threeptfn\ from \finalthrsmall\
using \expectval . So in order to compute three point functions
of the stress tensor in the hypothetical
three dimensional field theory corresponding
to a nearly $dS_4$ spacetime  all we need to do is
apply formula \expectval\ to our results in section four. To go to
the corresponding expectation values in $EAdS_4$ we just need to
multiply all $dS_4$ results by a minus sign which comes from
$R_{dS}^2 \to -R_{AdS}^2 $ and all correlators of the stress
tensor have such a factor in front in the tree level gravity
approximation.

Some of the points we explained above are specific to the
four dimensional $dS_4 $ case. The situation in $dS_5$ is rather
interesting. The computation of fluctuations for a massless scalar
field gives, outside the horizon,
\eqn\fluccal{
\langle f_{\vec k } f_{\vec k'} \rangle \sim  { H^3 } ( 2 \pi)^4
\delta( \vec k + \vec k') { 4 \over \pi} { 1 \over  k^4}
~,~~~~~~~~~~~~H =  R_{dS}^{-1}
}
On the other hand the wavefunction $\Psi \sim e^{ i S}$
 has the form
\eqn\wafn{
 iS = - { i \over 2} R_{ds}^3 \int {d^4k \over (2 \pi)^4 }
 f^0_{\vec k } f^0_{- \vec k}
 [  { k^2 \over 2 \eta^2_c} - { 1 \over 4} k^4 \log( - \eta_c k)  - i { \pi \over 8}
  k^4  + \alpha k^4
 ]
}
where $\alpha$ is a real number. Note that the only term contributing
to \fluccal\ is the real term proportional to $k^4$. All other
terms are purely imaginary.
From \wafn\  we can compute the non-local contribution to the
two point function
which gives
\eqn\strestwo{
\langle {\cal O} (\vec k) {\cal O} ( \vec k' ) \rangle_{dS_5} \sim
 (2 \pi)^4 \delta( \vec k + \vec k' ) i  R_{ds}^3   { 1 \over 4}  k^4 \log k
}
The $EAdS_5$ answer is given by the analytic continuation \analytic .
Notice that the $i$ is due to the fact that we have an odd
number of powers of $R_{dS}$ and is consistent with the fact that
the logarithmic term in the wavefunction is purely imaginary.
For the stress tensor this gives an imaginary central charge and imaginary
three point functions.
It is rather interesting that the two point function \fluccal\ is
related to a local term in the wavefunction, namely the term
proportional to $k^4$, which is the only real term. In other
words, the non-local piece in the wavefunction which
determines the stress tensor seems unrelated to the local
piece which determines the expectation value of the fluctuations.
In other words, $dS_5/CFT_4$ would tell us how to compute the non-local
piece in the wavefunction but will give us no information on the
local piece. On the other hand from the inflationary point of view
we would be  interested in computing \fluccal\ which depends on
the local part of the wavefunction, or the partition function of
the CFT. Maybe in dS/CFT we are only allowed to use imaginary
counterterms, then the field theory should be such that it allows
the computation of the finite real local parts in the effective
action. Note that the real term in \wafn\ arises in the analytic
continuation \analytic\ from the term in the $EAdS_5$ wavefunction
that is  proportional to $k^4 \log ( z_c k)  \to - { \pi \over 2} k^4
+ k^4 \log (- \eta_c k)$. So still, in some sense, the real part of
the wavefunction \wafn\ is intimately related to the non-local term
in the wavefunction.
It looks like this will be the situation in all odd
dimensional $dS$ spaces. The $AdS_3$ case studied in \stromingerds\
 seems special because there is no bulk propagating graviton.
Stress tensor correlators in dS/CFT were also studied in
\vijaycounterterms \otherds .

Now let us reexamine the three point functions of stress tensor
operators in the limit
that one of the momenta is much smaller than the other two.
We can then approximate the small  momentum by zero.
This zero momentum insertion of the stress tensor can
be viewed as coming from an infinitesimal  coordinate transformation.
So we then know that the three point function is going
to be given by the change  of the two point function by
this coordinate transformation.
For example, an  insertion of the trace of the stress tensor at
zero momentum  is equivalent
to performing a rescaling of the coordinates without rescaling the
mass scale of the theory. Then the
three point function will be given by the scale dependence of
the two point function.
In other words
\eqn\corrform{
\langle 2 T^i_i(0) { \cal O}(k) {\cal O}(k') \rangle =
-  k^i { \partial \over \partial k^i }
\langle { \cal O}(k) {\cal O}(k') \rangle
}
This is the reason why three point functions in this limit are
proportional to the tilt of the scalar and tensor spectra respectively,
see \simpleget\ \smallktwogr . There is a similar argument for
the insertion of the traceless part of the stress tensor at zero
momentum. Formula \corrform  is valid to all orders in slow roll.

Notice that in order to compute observable quantities from
dS/CFT we will need to square the wavefunction and integrate
over some range of values of the couplings and the metric of
the space where the CFT is defined. In other words, in order
to compute some physically interesting quantity it is not enough
to consider the CFT on a fixed 3-manifold but over a range
of three manifolds.
This is the reason that expectation values in
dS are not simply given by analytic continuation of the ones in
EAdS  \otherds\ even though the wavefunction and correlation functions of
operators are given by analytic continuation.\foot{ This analytic
continuation is very  clear for fields with $ 2 m R_{dS_d} < d$. For
fields with  mass above this bound
it is not so clear what the right prescription is. In this paper
we focus our attention on the easy case. }
This makes it clear that even if dS/CFT is
true there is no  causality problem, one is not fixing the
final state of the universe. One fixes it as an auxiliary step
in order
to compute the wavefunction but in order to compute probabilities we
need to sum over all final boundary conditions.
A slightly different integral
over  boundary conditions arises also
in the $EAdS$ context when we consider certain relevant operators
\kw , or double trace operators \wittendouble . In those cases this
integration is the same as a change in the boundary condition. Note
that this is {\it not}
 what happens in the $dS$ context since we have the
{\it square} of the wavefunction.
 One might conjecture that $dS$ expectation
values are given by two CFTs (one for $\Psi$ and one for $\Psi^*$) coupled
together in some fashion. Note that then it is not clear if we should
view the resulting object as a local field theory since in the
resulting object is not defined on a fixed manifold since in order
to compute expectation values we need to integrate over the three metric.
The two copies of the CFT that we are talking about
 arise just at the future
boundary, so these two copies are different than the two copies
talked about in \wittends \stromingerds \otherds \vijaycounterterms .
In global coordinates in addition we have the past boundary. Throughout
this paper we have ignored the past  boundary
since we focused on distances
larger than the Hubble scale but smaller than the total size of the spatial
slice. In the Hartle and Hawking
prescription for the wavefunction of the universe the past and future
parts of the wavefunctions are  complex conjugates of each
other since the total wavefunction is real \hh .
It is natural to suspect that
 these two pieces can be thought of as
 $\Psi$ and $\Psi^*$ in our discussion above.

{\bf Acknowledgments }

I would like to thank E. Witten for many discussions and initial
collaboration on these issues. I would also like to thank E.
Komatsu, D. Spergel and P. Steinhardt for  discussions. I also
thank E. Silverstein and S. Weinberg for pointing out some typos and for some
questions that to a clearer presentation. I also thank S. Weinberg for pointing
out an error in the previous version of section 3.1.

 This work was supported in part by DOE grant
DE-FG02-90ER40542.

\appendix{A}{ Second order change of variables between the two gauges}

In this appendix we work out explicitly the change of variables
to second order between the gauge \firstg\ and the gauge \secondg .
Let us denote by $\tilde t$ the time coordinate in \secondg ,
by $t$ the one in \firstg . The time reparametrization is
$\tilde t = t + T $. First let us find the first order change of
variables between the two gauges. It is easy to see that we do
not need to do any spatial reparametrization to first order and
all we need to do is a time reparametrization.
The value of $T_1$ at first order, as well as the relation between
variables is
\eqn\firstrepar{
 T_1= - { \varphi_1 \over \dot \phi} = { \zeta_1 \over \dot \rho} ~,~~~~~~
 \zeta_1 = - { \dot \rho \over \dot \phi} \varphi_1
}
where the subindex reminds us that it is a first order relation.

Now we work this out to second order.
In order to go from the gauge \secondg\ where $\varphi$ is not zero
to \firstg\ where $\varphi$ is zero we need to do a time
reparametrization determined by
the equation  $\phi(t + T(t)) + \varphi(t+T(t)) = \phi(t)$ which
gives, to second order,
\eqn\valueofTapp{
T = - { \varphi \over \dot \phi} - {1 \over 2} { \ddot \phi \varphi^2
\over \dot \phi^3} + { \dot \varphi  \varphi \over \dot \phi^2 }
}
Under this change of variables we find that the metric in \secondg\
becomes
\eqn\intermmetr{
h_{ij}^r = e^{ 2 \rho( t+ T) } ( \delta_{ij} + \tilde \gamma_{ij}(t)
+
{ 1 \over 2} \tilde \gamma_{il} \tilde \gamma_{lj}  +
\dot {\tilde \gamma}_{ij} T + N^i_\varphi \partial_j T
+ N^j_\varphi \partial_i T - e^{ -2 \rho} \partial_i T \partial_j T )
}
where we have set $N=1$ in some second order terms and $N^i_\varphi$
is given in \expres .
This metric $h^r_{ij}$ does not yet obey the gauge condition
in \firstg . The violation of the gauge condition is due to
some second order terms, since we already saw that at first order
we do not need to do a spatial reparametrization.
The terms responsible for this violation are the last four  terms
in \intermmetr .
In order to make it obey those gauge condition it is
necessary to do a spatial reparametrization where $\tilde x^i =
x^i + \epsilon^i(x,t) $, where $\epsilon^i$ is of second order.
The condition that $\epsilon^i$ should obey is that
\eqn\condition{
\delta \hat
h_{ij}^{r } + \partial_i \epsilon^j + \partial_j \epsilon^i =
2 \alpha \delta_{ij} +
\mu_{ij}  ~,~~~~~~~~ \partial_i \mu_{ij} =0~,~~~~~~~~
\mu_{ii} =0
}
Where $\delta \hat
 h_{ij}^r$ represents the last four  terms in \intermmetr .
In order to solve this equation for $\epsilon^i$ it is convenient
to separate $\epsilon^i = \partial_i \epsilon + \epsilon^i_t $
where $\partial_i \epsilon_t^i =0$.
Taking the trace and $\partial_i \partial_j$ of \condition\
we obtain
\eqn\valueofalpha{\eqalign{
4 \alpha =&  \delta \hat
h_{ii}^{r } - \partial^{-2} \partial_i \partial_j
 \delta \hat h_{ij}^{r }
\cr
=& -  \partial_i T \partial_i T e^{ -2 \rho} + 2 N^i_\varphi
\partial_i T  + \partial^{-2}
 \partial_i \partial_j  (\partial_i T \partial_j T) e^{ -2 \rho}
\cr
& ~~- 2 \partial^{-2} \partial_i \partial_j ( N^i_\varphi \partial_j T )
- \partial^2 \dot {\tilde \gamma}_{ij} \partial_i \partial_j T }}
We can similarly compute what $\epsilon^i$ are and then
find $\mu_{ij}$ from \condition .
It turns out that for our purposes we can write
\eqn\exprformu{
\mu_{ij} = \delta \hat h_{ij}^{r} + {\rm rest}
}
where the last terms are given by all other  terms in \condition . These
terms
vanish when integrated against a function which is traceless and
divergenceless, which means that these extra terms do not contribute
to our computation.
In conclusion, after making this spatial reparametrization we
can put the metric in a form such that it obeys the gauge
\firstg\ to second order.

We find then that the final field redefinition is given by
\eqn\finalf{\eqalign{
\zeta = & \rho(t+ T) - \rho(t) + \alpha
\cr
\gamma_{ij} & = \tilde \gamma_{ij} + \dot {\tilde \gamma}_{ij} T +
\mu_{ij}
}}
where $T$, $\alpha$ and $\mu_{ij}$ are given above.

It is convenient to define a variable $\zeta_n$ through the relation
\rela . Here we are thinking of $\zeta_n$ as a convenient parameterization
of the variable $\varphi$.
Then the explicit change of variables between the two gauges is
\eqn\fredexpl{\eqalign{
\zeta = & \zeta_n +
 { 1 \over 2}
{ \ddot  \phi \over \dot \phi \dot \rho}
\zeta^2_n + { 1 \over 4} { \dot \phi^2 \over \dot \rho^2 }
\zeta^2_n
+
\cr ~&~
+ { 1 \over \dot \rho} \dot \zeta_n \zeta_n
-{1 \over 4} { e^{ -2 \rho} \over \dot \rho^2 } (\partial \zeta_n)^2
+ { 1\over 4}
{ e^{ -2 \rho} \over \dot \rho^2 } \partial^{-2} \partial_i
\partial_j ( \partial_i \zeta_n \partial_j \zeta_n)
+ {1 \over 2} { 1 \over \dot \rho} \partial_i \chi_n \partial_i \zeta_n
\cr
 &~ - { 1 \over 2} { 1\over \dot \rho} \partial^{-2} \partial_i
\partial_j ( \partial_i
 \chi_n \partial_j \zeta_n)
- { 1 \over 4}{ 1 \over \dot \rho}
 \dot {\tilde \gamma}_{ij} \partial_i \partial_j \zeta_n
\cr
\gamma_{ij} = & \tilde \gamma_{ij} +
\cr
  & ~+ { 1 \over \dot \rho} \dot {\tilde
\gamma}_{ij} \zeta_n  - { e^{ - 2 \rho} \over
\dot \rho^2} \partial_i \zeta_n \partial_j \zeta_n +
{ 1\over \dot \rho} (\partial_i \chi_n \partial_j \zeta_n +
\partial_j \chi \partial_i \zeta_n )
}}
we see that only the first line of each field redefinition is non-vanishing
outside the horizon. So fortunately we do not need to take all
these terms into account in the computations in the paper.
We did check however that the terms proportional to the equations
of motion that arise when we integrate by parts the lagrangian in
the gauge \firstg\ in order to make it look more like the
lagrangian in \secondg\ are indeed precisely the ones that
lead to the above field redefinitions.

\listrefs

\bye